\documentclass[12pt]{article}
\usepackage{amsmath}
\usepackage{amsfonts}
\usepackage{amssymb}
\usepackage{mathtools}
\usepackage{enumitem}
\usepackage{}
\usepackage{graphicx}
\usepackage{color}
\usepackage[round]{natbib}
\usepackage{titlesec}
\usepackage{amsthm}
\usepackage{tabularx}
\usepackage{booktabs}
\usepackage{multirow}
\usepackage{colortbl}
\usepackage{arydshln}
\usepackage{titling}
\usepackage{tikz}
\usetikzlibrary{fit,shapes.misc}

\newcommand{\blind}{1}
\titlelabel{\thetitle. }
\titleformat*{\section}{\Large\bfseries}
\graphicspath{{./fig/}}
\setcitestyle{aysep={}}
\begin{document}
\def\spacingset#1{\renewcommand{\baselinestretch}%
{#1}\small\normalsize} \spacingset{1}


\if1\blind
{
  \title{\vspace{-50pt}\bf The More Data, the Better? Demystifying Deletion-Based Methods in Linear Regression with Missing Data}
  \author{Tianchen Xu\\
  Mailman School of Public Health, Columbia University,\\ New York, NY, 10032 \\tx2155@columbia.edu  
 \\ and \\ Kun Chen\\
  Department of Statistics, University of Connecticut, Storrs, CT 06269 \\ and \\ Gen Li \\
  School of Public Health, University of Michigan, Ann Arbor, MI 48109}
\maketitle
} \fi

\if0\blind
{
  \bigskip
  \bigskip
  \bigskip
  \begin{center}
    \spacingset{1.5}{\LARGE\bf The More Data, the Better? Demystifying Deletion-Based Methods in Linear Regression with Missing Data}
\end{center}
  \medskip
} \fi

\begin{abstract}
We compare two deletion-based methods for dealing with the problem of missing observations in linear regression analysis. One is the complete-case analysis (CC, or listwise deletion) that discards all incomplete observations and only uses common samples for ordinary least-squares estimation. The other is the available-case analysis (AC, or pairwise deletion) that utilizes all available data to estimate the covariance matrices and applies these matrices to construct the normal equation. We show that the estimates from both methods are asymptotically unbiased and further compare their asymptotic variances in some typical situations. 
Surprisingly, using more data (i.e., AC) does not necessarily lead to better asymptotic efficiency in many scenarios. Missing patterns, covariance structure and true regression coefficient values all play a role in determining which is better. 
We further conduct simulation studies to corroborate the findings and demystify what has been missed or misinterpreted in the literature. Some detailed proofs and simulation results are available in the online supplemental materials.
\end{abstract}

\noindent%
{\it Keywords:} asymptotic variance; available-case analysis; complete-case analysis; missing data.
\vfill

\newcommand{\sx}{\boldsymbol{\Sigma}_x}
\newcommand{\sz}{\boldsymbol{\Sigma}_z}
\newcommand{\sxy}{\boldsymbol{\Sigma}_{xy}}
\newcommand{\sy}{\boldsymbol{\Sigma}_{y}}
\newcommand{\Sx}{\boldsymbol{S}_x}
\newcommand{\Sxy}{\boldsymbol{S}_{xy}}
\newcommand{\Sy}{\boldsymbol{S}_{y}}
\newcommand{\Yb}{\boldsymbol{Y}}
\newcommand{\oneb}{\boldsymbol{1}}
\newcommand{\Xb}{\boldsymbol{X}}
\newcommand{\Zb}{\boldsymbol{Z}}
\newcommand{\eb}{\boldsymbol{\varepsilon}}
\newcommand{\bb}{\boldsymbol{\beta}}
\newcommand{\cov}{\operatorname{Cov}}
\newcommand{\Ib}{\operatorname{I}}
\renewcommand{\vec}{\operatorname{Vec}}
\newcommand{\muxb}{\boldsymbol{\mu}_x}
\newcommand{\mub}{\boldsymbol{\mu}}
\renewcommand{\sb}{\boldsymbol{\Sigma}}
\newcommand{\Sb}{\boldsymbol{S}}
\newcommand{\Qb}{\boldsymbol{Q}}
\newcommand{\Phib}{\boldsymbol{\Phi}}
\newcommand{\Rb}{\boldsymbol{R}}
\newcommand{\rhob}{\boldsymbol{\rho}}
\newcommand{\alphab}{\boldsymbol{\alpha}}
\newcommand{\gammab}{\boldsymbol{\gamma}}
\def\gop{{\buildrel P \over \longrightarrow}}
\def\god{{\buildrel {d} \over \longrightarrow}}
\def\goas{{\buildrel {\rm a.s.} \over \longrightarrow}}\
\newcommand{\Deltab}{\boldsymbol{\Delta}}
\newcommand{\e}{\operatorname{E}}
\newcommand{\var}{\operatorname{Var}}
\newtheoremstyle{prop1}
{3pt}
{3pt}
{}
{}
{\bf}
{:}
{.5em}
{\thmname{#1}\thmnumber{ #2}\thmnote{ (#3)}}
\theoremstyle{prop1}
\newtheorem{prop}{Proposition}[section]

\newpage
\spacingset{1.7} 
\section{INTRODUCTION}
Missing data are very common in linear regression analysis. \citet{dong2013principled} described missing data as ``a rule rather than an exception in quantitative research.''
For instance, longitudinal data may be incomplete due to unexpected dropout, and survey data may be incomplete due to refusal of respondents or wrong answers. Since inappropriate treatments on missing data can severely undermine the validity of inference and conclusion of a study, researchers have developed many methods to conquer this challenge. 

Deletion-based methods that usually involve complete-case analysis (CC, or listwise deletion) and available-case analysis (AC, or pairwise deletion) are the simplest and most frequently used for dealing with missing data \citep{dong2013principled}. For instance, \citet{peng2006advances} examined $569$ papers with missing data published in $11$ education journals from $1998$ to $2004$ and found that $552$ $(97\%)$ employed deletion-based methods; \citet{lang2018principled} reviewed $169$ papers with missing data in \textit{Prevention Science} from February 2013 to July 2015 and found that $62$ $(37\%)$ studies used deletion-based methods. Especially recently, there is an increasing tread on applying the AC method or its variants to high dimensional data such as block-missing multi-modality datasets where each subject has missing blocks from certain modality sources \citep{yu2020optimal, xue2020integrating}. 
The CC method utilizes the complete dataset in which any incomplete rows are discarded and is the default setting for many multivariate procedures and regressions analysis in popular statistical packages such as SAS, SPSS, SYSTAT and R. The AC method computes statistics using the rows for which every constituent variables are observed and is the default setting for descriptive, correlation, and regression analysis when using either correlation or covariance matrices in SAS, SPSS and SYSTAT. The \verb!cov! function and \verb!regtools! package in R also provide AC analysis for correlation estimation and linear regression. 
The goal of this article is to compare the performance of these two methods. Particularly, we mainly focus on the classical low-dimensional settings with the assumption that the proportion of complete observation is positive (to ensure the CC method is feasible) and some typical block-wise missing patterns.  

Other mainstream treatments for missing data in regression analysis include: 1) imputation, 2) weighting, and 3) maximum-likelihood based methods \citep{lang2018principled, little2019statistical}. 
Imputation methods try to impute the missing part of the dataset. Single imputation often imputes the missing values with some fixed values (e.g., mean values), random drawn values from the same variable (simple hot-deck) or predictive values from other variables. Multiple imputation (MI) imputes the missing data while acknowledging the uncertainty associated with the imputed values \citep{rubin1977formalizing, rubin1996multiple}. Weighting approaches discard incomplete samples and assign a new weight to each subject according to some missing features to reduce the bias and variance of the final inference \citep{seaman2013review}. Maximum-likelihood methods such as information maximum likelihood (FIML, also known as direct maximum likelihood) consider only the observed samples when calculating the sample log-likelihood function and maximize it using EM algorithm to estimate parameters \citep{enders2001relative, olinsky2003comparative}. Generally speaking, no technique is universally better than others. 
Under the missing completely at random (MCAR) assumption, deletion-based methods are the only fully automatic methods, while other methods typically require specific modeling and careful tuning by users \citep{calzolari1987computational, sinharay2001use, enders2001relative, enders2008note, hardt2012auxiliary, seaman2013review}, and some of these can be unstable when a complete block of the data is missing \citep{yu2020optimal}.


Although deletion-based methods are popular due to its simplicity, there is no consensus about whether the AC or the CC is better. There have been intense debates in the literature about the merits and flaws of different deletion-based methods. In particular, AC and CC are often in the center of the controversy. \citet{glasser1964linear} is the first researcher (as far as we know) who systematically introduced the AC estimator in the context of linear regression. He argued that the AC estimator is consistent and derived its asymptotic variance. In the simulation study with two predictors ($p=2$), \citet{glasser1964linear} concluded that the AC estimator is in general better that the CC estimator if the correlation between two predictors is less than $0.58$. However, \citet{haitovsky1968missing} pointed out that \citet{glasser1964linear}'s asymptotic result which does not involve the true regression coefficients ($\bb$) was not accurate and provided the right asymptotic covariance. He also reached an opposite conclusion that ``listwise deletion (the CC estimator) is judged superior in almost all the cases'' by considering nine simulation scenarios. In contrast to \citet{haitovsky1968missing}'s findings, \citet{kim1977treatment} did another simulation study and claimed their setting is more typical in sociological studies. The result indicated that the AC method performs better than the CC estimator by using the correlation structure among predictors in \citet{blau1967american}'s book. In the following decades, these contradictory papers were frequently cited by researchers to show the comparison between two methods are not fully settled \citep{little1992regression, allison2001missing, pigott2001review}. 


The rest of the paper is organized as follows. In Section~\ref{sec:backgroud}, we review the existing results of both methods. In Section~\ref{sec:comparison}, we compare the performance of any scalar regression coefficient estimator in realistic situations. We show that the estimators from both methods are asymptotically unbiased and using more data (i.e., AC) does not necessarily lead to better asymptotic performance. It is necessary to look in the missing patterns, covariance structure and true regression coefficients together to determine which method is better.  In Section~\ref{sec:simulation}, we conduct simulation studies based on \citet{kim1977treatment}'s settings to verify our theoretical propositions and validate our findings in Section~\ref{sec:comparison}. With the guidance of the theoretical results, we are able to find out what was missed or misinterpreted in the previous work and provide our suggestions. In the last section, we discuss further research directions.

\section{BACKGROUND}\label{sec:backgroud}
\subsection{Asymptotic Results for Complete Case}\label{sec:complete}
Let $\Xb=(X_1, X_2,\cdots X_p)^T\in\mathbb{R}^p$ be a random vector. Let $Y\in\mathbb{R}$ be a random variable such that
\[
Y=\Xb^T\bb+\varepsilon,
\]
where $\boldsymbol{\beta}=(\beta_1,\cdots,\beta_p)^T\in\mathbb{R}^{p}$ is a coefficient vector, $\varepsilon\in\mathbb{R}$ is a random variable with mean $0$ and variance $\sigma^2$. Furthermore, we assume $X_j$ is independent of $\varepsilon$. 

Let $\Zb=(Z_1,\cdots,Z_{p+1})^T\triangleq\left(\Xb^T, Y\right)^T$ to be a $(p+1)$-dimensional random vector with mean $\mub$ and non-singular covariance matrix $\sb$. Assume all fourth-order moments of $\Zb$ are finite. 
Partition $\sb$ conformably as follows:
\begin{gather*}
    \sb = \begin{pmatrix}
        \sx& \sxy\\
        \sxy^T & \sy
    \end{pmatrix},
\end{gather*}
where $\sx = \cov(\Xb)$, $\sxy = \cov(\Xb,Y)=\sx \bb$, $\sy = \var (Y)=\bb^T \sx \bb+\sigma^2$. 
Let $\mu_j$ denote the $j$th element in~$\mub$. Let $\sigma_{jk}$ denote the $(j,k)$th element in $\sb$, and conventionally we use $\sigma^2_j$ to denote the elements on the diagonal of $\sb$ (i.e., $\sigma^2_j = \sigma_{jj}$). 

We collect a set of observation data $\{Z_{1i},\cdots Z_{p+1,i}\}_{i=1,\cdots,n}$ from $n$ independent samples and assume there are not any missing data in this section. Define the sample covariance matrix $\Sb=[s_{jk}]$ with entries:
\begin{align*}
    s_{jk} = \frac{1}{n}\sum_{i=1}^n (Z_{ji}-\bar Z_j)(Z_{ki}-\bar Z_k),
\end{align*}
where $\bar Z_j$ is the sample mean of $Z_j$ ($j=1,\cdots,p+1$). Similar to $\sb$, we also partition $\Sb$ into four parts correspondingly:
\begin{align*}
    \Sb =  \begin{pmatrix}
        \Sx& \Sxy\\
        \Sxy^T & S_y
        \end{pmatrix},
\end{align*}
where $\Sx$, $\Sxy$, $S_y$ are the sample covariance/variance of $\Xb$, $(\Xb, Y)$ and $Y$.
Then the least-squares estimator of $\bb$ is well known:
\begin{align*}
    \hat \bb = \Sx^{-1}\Sxy.
\end{align*}

The sample quantities $\Sb$ and $\hat \bb$ are consistent estimators of their theoretical counterparts $\sb$, $\bb$ respectively and are asymptotically normally distributed \citep{rao1973linear}. The former is guaranteed by the Lindeberg-Levy Central Limit Theorem and the latter can be derived from the Delta method. 
\begin{prop}[\citet{rao1973linear}]
Let $\Sb$ be the sample covariance matrix of r.v $\Zb=\left(\Xb^T, Y\right)^T$, then 
\begin{align*}
    \sqrt{n}(\operatorname{vec}(\Sb)-\operatorname{vec}(\sb))\god N\left(\boldsymbol{0},\Phib\right),
\end{align*}
where the asymptotic covariance $\Phib$ consists of elements $\phi_{(ij)(mn)}\triangleq\cov(s_{jk},s_{mn})$:
\begin{align*}
    \phi_{(jk)(mn)} = \e(Z_j-\mu_j)(Z_k-\mu_k)(Z_m-\mu_m)(Z_n-\mu_n)-\sigma_{jk}\sigma_{mn}.
\end{align*}
\end{prop}
\begin{prop}[\citet{rao1973linear}]
    Let $\hat \bb$ be the least-squares estimator of $\bb$ in the aforementioned regression, then 
    \begin{align*}
        \sqrt{n}(\hat \bb - \bb)\god N_p(\boldsymbol{0}, \Deltab \Phib \Deltab^T),
    \end{align*}
where $\Deltab$ denotes the matrix of partial derivatives of function $\hat\bb(\Sb)$ evaluated in $\sb$. 

The form of $\Deltab$ and $\Phib$ depends on the way of vectorizing $\Sb$. In Appendix A (online supplemental material), we provide an example of vectorizing $\Sb$ in columns (i.e., $\operatorname{vec}(\sb)$). Similar results have been obtained in the literature; see \citet{white1980using, VANPRAAG1981139, bentler1985efficient} for example.
\end{prop}

\subsection{Asymptotic Results for Incomplete Case}
Suppose there are missing values in predictor matrix $\Xb$. Following \citet{little1982models}, the missing pattern is independent of the values of predictors (i.e., missing completely at random, MCAR). Let $\Rb=[R_{ji}]$ ($j=1,\cdots,p+1; i=1,\cdots,n$) be an indicator matrix that
\begin{align*}
    R_{ji}=\begin{dcases}
        0 & \text{if } Z_{ji} \text{ is not observed},\\
        1 & \text{if } Z_{ji} \text{ is observed}.\\
    \end{dcases}
\end{align*}
\subsubsection{Available-Case Analysis}
``Available-case analysis (AC) tries to use the largest possible sets of available cases to estimate individual parameters'' \citep{little1992regression, pigott2001review}. 
Define the sample covariance matrix in the AC method $\Sb_{AC}=[s^{AC}_{jk}]$ with entries:
\begin{align*}
    s^{AC}_{jk} = \frac{1}{n_{jk}}\sum_{i\in \tau_{jk}} \Big(Z_{ji}-\frac{1}{n_{jk}}\sum_{l\in \tau_{jk}}Z_{jl}\Big)\Big(Z_{ki}-\frac{1}{n_{jk}}\sum_{l\in \tau_{jk}}Z_{kl}\Big),
\end{align*}
where $\tau_{jk}=\{i:R_{ji}R_{ki}=1\}$ is the index set of samples that both $Z_j$ and $Z_k$ are observed; $n_{jk}$ is the size of $\tau_{jk}$ (i.e., $n_{jk}=\sum_{i=1}^n R_{ji}R_{ki}$). A defect of the AC method is that the estimated covariance matrix $\Sb_{AC}$ might not be positive definite. However, \citet{van1985least} pointed out that the probability of $\Sb_{AC}$ being positive definite tends to $1$ as the sample size increases. Similar to $\Sb$, we partition $\Sb_{AC}$ into $\Sx^{AC}$, $\Sxy^{AC}$, $S_y^{AC}$ and define the AC estimator $\hat \bb_{AC}$ as follows:
\begin{align*}
    \hat \bb_{AC} = \left\{\Sx^{AC}\right\}^{-1}\Sxy^{AC}.
\end{align*}

Let $q_{j}$ be the proportion of the cases with $Z_j$ observed (i.e., $q_j=\frac{1}{n}\sum_{i=1}^n R_{ji}$), and $q_{jk}$ be the proportion of the cases with both $Z_j$ and $Z_k$ observed (i.e., $q_{jk}=\frac{1}{n}\sum_{i=1}^n R_{ji}R_{ki}$). Similarly, we also define $q_{jkm}$ and $q_{jkmn}$. For the AC estimator, the following proposition holds: 
\begin{prop}[\citet{van1985least}]
Under the MCAR assumption, assuming that the observing proportions (i.e., $q_j,q_{jk},q_{jkm},q_{jkmn}$) are not zero and remain the same as sample size $n$ goes to infinity, the asymptotic distribution of $\hat\bb_{AC}$ is given by: 
\begin{align*}
    \sqrt{n}(\hat\bb_{AC}-\bb)\god N_p(\boldsymbol{0}, \Deltab (\Phib\circ \Qb) \Deltab^T),
\end{align*}
where  $\Qb$ consists of elements $q_{(jk)(mn)}=\frac{q_{jkmn}}{q_{jk}q_{mn}}$ corresponding to $\phi_{(jk)(mn)}$; $\circ$ represents the Hadamard product.

From the proportion, we conclude that $\bb_{AC}$ is asymptotically unbiased and its asymptotic variance is $\Deltab (\Phib\circ \Qb) \Deltab^T/n$, obtained by multiplying a specific factor $q_{(jk)(mn)}$ to $\phi_{(jk)(mn)}$ in $\Phib$ that is from the variance of $\hat\bb$ in the complete case.

\end{prop}

\subsubsection{Complete-Case Analysis}
Complete-case analysis (CC) only utilizes the complete samples without any missing data and usually serves as a baseline for comparisons. The CC estimator $\hat\bb_{CC}$ is exactly the same as $\hat\bb$ in Section~\ref{sec:complete} except that the dataset is constrained to complete samples. Therefore, the CC method is feasible only when there exist sufficient number of complete cases. 
\begin{prop}
    Let $\tilde q$ denote the proportion of samples that have complete observations, and assume $\tilde q> 0$ is a constant. Under the MCAR assumption, the CC estimator $\hat \bb_{CC}$ follows:
\begin{align*}
    \sqrt{n}(\hat\bb_{CC}-\bb)\god N_p(\boldsymbol{0}, \Deltab \Phib \Deltab^T/\tilde q),
\end{align*}

\end{prop}
Similar with the AC estimator, $\hat \bb_{CC}$ is also asymptotically unbiased and its asymptotic variance is $\Deltab \Phib \Deltab^T/(n\tilde q)$.

\section{COMPARISON BETWEEN AC AND CC}\label{sec:comparison}
Somewhat surprisingly, although AC makes better use of data by accounting for all available data points, many simulation studies show that AC is markedly inferior to CC on highly correlated data and can be superior to CC on weakly correlated data \citep{haitovsky1968missing, kim1977treatment, little1989analysis}. Since both $\hat\bb_{AC}$ and $\hat\bb_{CC}$ are consistent estimators of $\bb$, we compare their asymptotic variances in the article. Let $V_{CC}$, $V_{AC}$ denote the asymptotic variance of $\hat\bb_{CC}$, $\hat\bb_{AC}$ respectively. Then the difference $V_{D}$ is:
\begin{align*}
   V_{D}& = V_{CC}-V_{AC}\\
    &=\frac{1}{n\tilde q}\Deltab \left\{\Phib\circ (1-\Qb \tilde q)\right\}\Deltab^T.
\end{align*}
Neither method is uniformly better than the other with any fixed missing pattern (see more detailed explanation in Appendix B (online supplemental material)). It turns out that we have to look into the covariance structure $\sb$, true coefficient $\bb$ together with missing pattern $\Qb$ to determine which method is better. 

\subsection{Asymptotic Variance of Estimating an Individual Coefficient}
Comparing asymptotic covariance matrices of all coefficients is rather complicated. We can gain insights by focusing on comparing the variance of estimating an individual coefficient using either the AC or the CC method. This is a relevant task in many real applications. For example, in genetics, we often want to test the association of a disease and a genetic locus while adjusting for additional clinical covariates. Here we assume $\Zb$ follows an elliptical distribution and obtain the asymptotic variance of $\hat\beta_1$ in both methods without loss of generality. Under the MCAR and elliptical distribution assumption, the asymptotic variance of $\hat\beta_1$ is as follows:
\begin{gather}
    n\cdot V_{AC,\hat \beta_1} = (1+\kappa)\left\{ \sum_{g=1}^p c_g \beta_g^2 + \sum_{g=1}^p\sum_{h=g+1}^p d_{gh}\beta_g\beta_h + \left(\sum_{j=1}^p \frac{r_{1j}^2 \sigma_j^2 }{q_j} + \sum_{j=1}^p\sum_{k=j+1}^p 2 r_{1j}r_{1k}\sigma_{jk}\frac{q_{jk}}{q_jq_k}\right)\sigma^2\right\},\label{eqn:singleac} \\
    n\cdot V_{CC, \hat \beta_1} =(1+\kappa) \left(\sum_{j=1}^p r_{1j}^2 \sigma_j^2  + \sum_{j=1}^p\sum_{k=j+1}^p 2 r_{1j}r_{1k}\sigma_{jk}\right)\frac{\sigma^2}{\tilde q} = (1+\kappa)\frac{r_{11}\sigma^2}{\tilde q},\label{eqn:singlecc}
 \end{gather}
where $c_g, d_{gh}$ are in Appendix C (online supplemental material); $r_{jk}$ is the $(j,k)$th element in $\sb^{-1}$ (e.g., $r_{1j}$ is the $j$th element in the first row of $\sb^{-1}$). We also notice when all proportions (i.e., $q_j, q_{jk}$, etc) are equal, namely there is no mismatched observations, then $c_g=d_{gh}=0$ and the variance of the AC estimator coincides with that of the CC estimator as expected. 

\textbf{Remark:} The reason for assuming an elliptical distribution of $\Zb$ is to simplify the fourth central moments involved in $V_D$. A special case is to assume $\Zb$ follow a multivariate normal distribution for which the fourth central moments can be expressed in terms of its covariance matrix by Isserlis' theorem \citep{isserlis1918formula}. In this article, we adopt a more general assumption that $\Zb$ follows an elliptically contoured distribution \citep{owen1983class} that includes not only  multivariate normal distribution, but also fatter-tailed distributions such as multivariate $t$-distribution, multivariate logistic distribution, and thinner-tailed distributions such as sub-Gaussian $\alpha$-stable distribution.
\citet{bentler1983some} introduced a kurtosis parameter $\kappa$ to link the fourth moments with the covariance matrix:
\begin{align*}
    \e(Z_j-\mu_j)(Z_k-\mu_k)(Z_m-\mu_m)(Z_n-\mu_n) = (1+\kappa)(\sigma_{jk}\sigma_{mn}+\sigma_{jm}\sigma_{kn}+\sigma_{jn}\sigma_{km})
\end{align*}
where $\kappa = \frac{\e(Z_j-\mu_j)^2}{3\{\e(Z_j-\mu_j)^2\}^2}-1$ is one-third of the excess kurtosis for each marginal r.v $Z_j$. In our regression setting, $\kappa$ is always larger than $-1/2$ \citep{bentler1986greatest}. For normal distribution, $\kappa=0$. 
There are several ways to estimate the common kurtosis parameter from the data (See Appendix D (online supplemental material)). 

\subsection{Comparison of $\text{\bf Var}\boldsymbol{(\hat \beta_1)}$ under Special Missing Patterns}\label{sec:comparison_single}
\begin{figure}[!ht]
    \textbf{(a)}\hspace{0.47\textwidth}\textbf{(b)}\\
    {\centering
    \includegraphics[width=.99\textwidth]{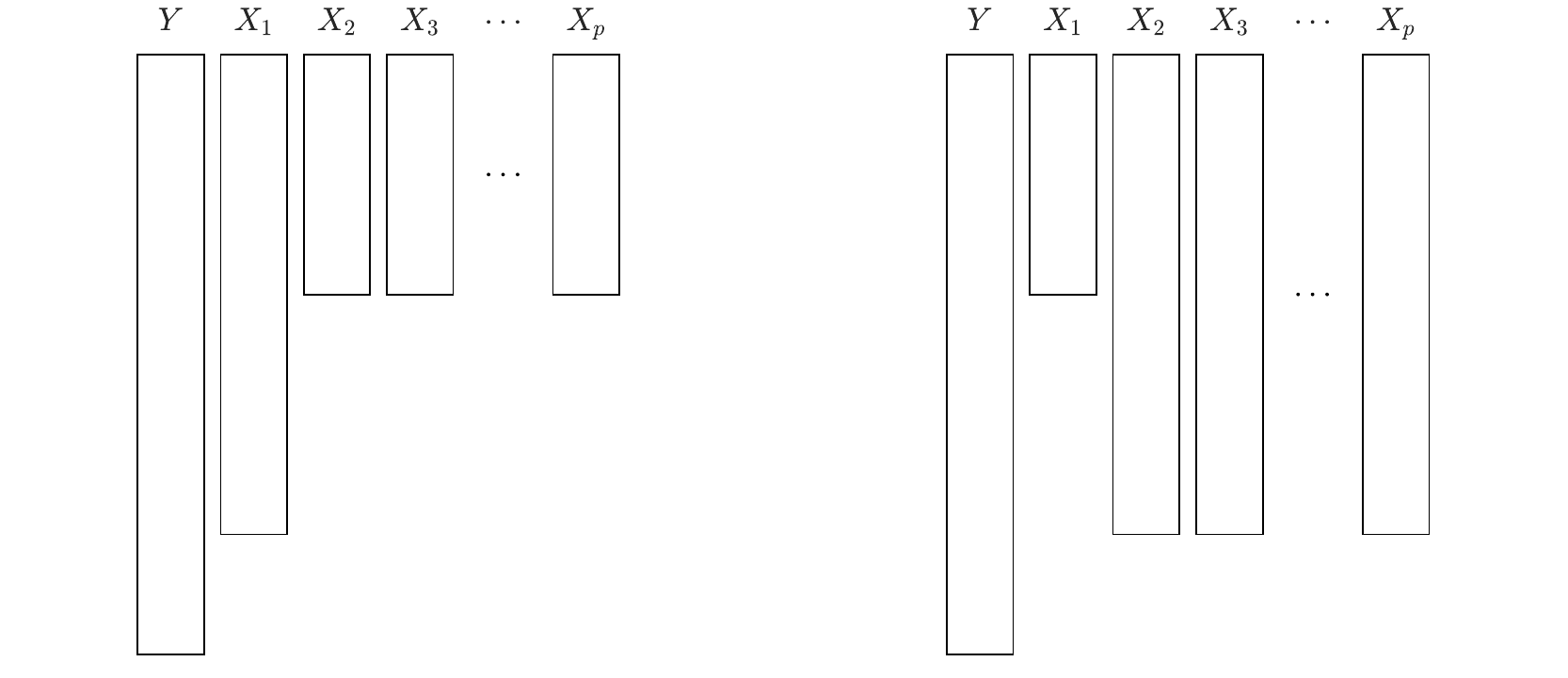}}
    \caption{Illustration of two missing patterns.
    \label{fig:missing}}
\end{figure}

As we can see from expressions~\eqref{eqn:singleac}, \eqref{eqn:singlecc}, a very general missing pattern results in a complex formula. In this section, we assume $X_2$ to $X_p$ follow the same missing pattern and explore the asymptotic variance of $\hat\beta_1$ in both methods. As shown in Figure~\ref{fig:missing}, we focus on two missing patterns. The pattern (a) is a \textit{unit monotone missing pattern} and the pattern (b) is \textit{univariate missing pattern} if predictors $X_2$ to $X_p$ are complete \citep{little1992regression}. 


\subsubsection{Missing Pattern (a)}\label{sec:mono}
Consider the unit monotone missing pattern (a) shown in Figure~\ref{fig:missing}(a). Let $q_1$ denote the observed proportion of $X_1$; $q_{-1}$ be the observed proportion of $X_j\,(j\ge 2)$. In addition, we assume available samples in $X_2$ to $X_p$ is a subset of $X_1\,(q_1> q_{-1})$  to have the monotone missing. 

According to expressions~\eqref{eqn:singleac} and \eqref{eqn:singlecc}, we obtain the asymptotic variance of $\hat \beta_1$ in both methods and calculate the difference $V_{D,\hat\beta_1}$.
Let $f(\bb)\triangleq n\cdot V_{D,\hat\beta_1}$ denote the difference of asymptotic variance as a function of $\bb$:
\begin{align*}
    f(\bb)=& -r_{11}^2\left(\frac{1}{q_{-1}}-\frac{1}{q_1}\right) \sum_{g=2}^p(\sigma_{1g}^2+2\kappa \sigma_{1g}^2+\sigma_1^2\sigma_{g}^2+\kappa \sigma_1^2\sigma_g^2)\beta_g^2\\
    &\qquad - 2r_{11}^2\left( \frac{1}{q_{-1}}-\frac{1}{q_1} \right) \sum_{g=2}^p\sum_{h=g+1}^p(\sigma_{1g}\sigma_{1h}+2\kappa\sigma_{1g}\sigma_{1h}+\sigma_1^2\sigma_{gh}+\kappa\sigma_1^2\sigma_{gh}) \beta_g\beta_h \\
    & \qquad+ (1+\kappa)\left(\frac{1}{q_{-1}}-\frac{1}{q_1}\right)\left(2r_{11}-r_{11}^2\sigma_1^2\right)\sigma^2.
\end{align*}
The true coefficient $\beta_1$ is not involved in this expression. If $f(\bb)>0$, then the AC estimator is better. 

We find that $\left(\frac{1}{q_{-1}}-\frac{1}{q_1}\right)$ is a key quantity in $f(\bb)$. NO matter which method is better, when we fix all other parameters, the larger the difference between $1/q_{-1}$ and $1/q_1$, the larger the difference of the two methods.

A special case is that all predictors are independent:
\begin{align*}
    f(\bb)=& \frac{1}{\sigma_1^2}(1+\kappa)\left(\frac{1}{q_{-1}}-\frac{1}{q_1}\right)\left(\sigma^2- \sum_{g=2}^p\sigma_g^2\beta_g^2\right).
\end{align*}
The AC estimator is better when $f(\bb)>0$, so we have the following proposition:

\begin{prop}
    In missing pattern (a), assuming all predictors are independent, the AC estimator is asymptotically better if and only if:
    \begin{align*}
        \sum_{g=2}^p\sigma_g^2\beta_g^2<\sigma^2. 
    \end{align*}
\end{prop}
We can rewrite the inequality as $\sum_{g=2}^p\left(\frac{\sigma_g\beta_g}{\sigma}\right)^2<1$, which means when the sum of squares of the standardized coefficients (except for $X_1$) are less than 1, the AC estimator of $\beta_1$ is better. 

For the general case that predictors are not independent, we further discuss the behavior of $f(\bb)$ under two scenarios where $p=2$ or $p\ge 3$. \\

\noindent\textbf{Scenario 1, $\bf p=2$:}\\
In this scenario, we only have predictors $X_1$, $X_2$ in our model. Then $f(\bb)$ is simplified as:
\begin{align*}
    f(\beta_2)=& -r_{11}^2(\sigma_{12}^2+2\kappa \sigma_{12}^2+\sigma_1^2\sigma_{2}^2+\kappa \sigma_1^2\sigma_2^2)\left(\frac{1}{q_{-1}}-\frac{1}{q_1}\right) \beta_2^2\\
    & \qquad+ (1+\kappa)\left(\frac{1}{q_{-1}}-\frac{1}{q_1}\right)\left(2r_{11}-r_{11}^2\sigma_1^2\right)\sigma^2.
\end{align*}

It is obvious that when the constant term (that does not involve $\beta_2$) is negative, $f(\beta_2)$ is always less than $0$ (i.e., CC is better). Therefore, we have the following proposition:
\begin{prop}
(See Appendix E (online supplemental material) for proof) In missing pattern (a) with two predictors, a sufficient condition that the CC estimator of $\beta_1$ is asymptotically better than the AC is:
\begin{align*}
    \frac{\sigma_1^2\sigma_2^2}{\sigma_{12}^2}<2.
\end{align*}
\end{prop}
This proposition shows that if the correlation between two predictors is  strong (i.e., $|\rho_{12}|>\frac{\sqrt{2}}{2}$), the AC estimator is always worse.

\begin{figure}[!hb]
    \textbf{(a)}\hspace{0.47\textwidth}\textbf{(b)}\\
    {    \centering
    \includegraphics[width=.49\textwidth]{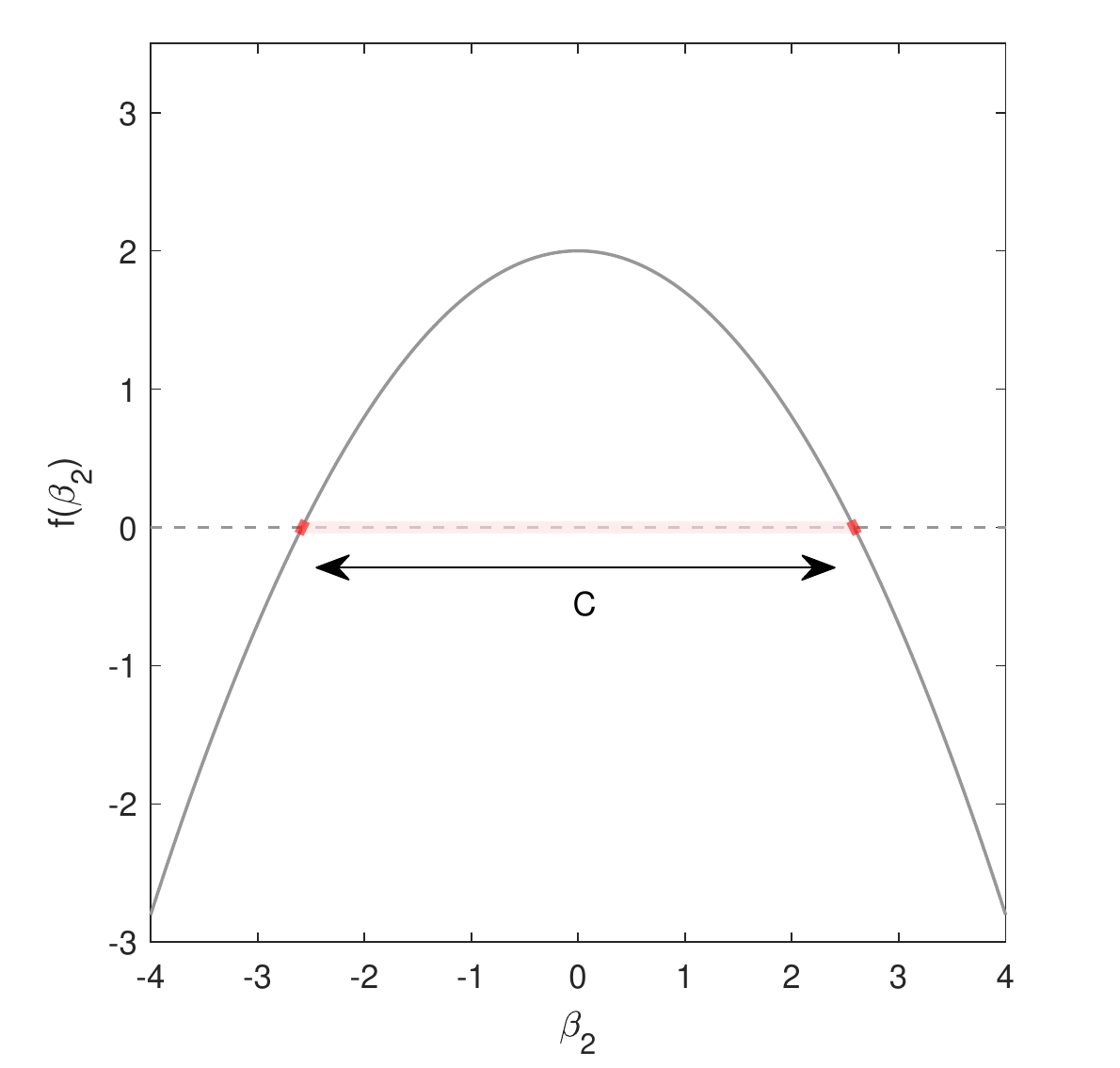}
    \includegraphics[width=.49\textwidth]{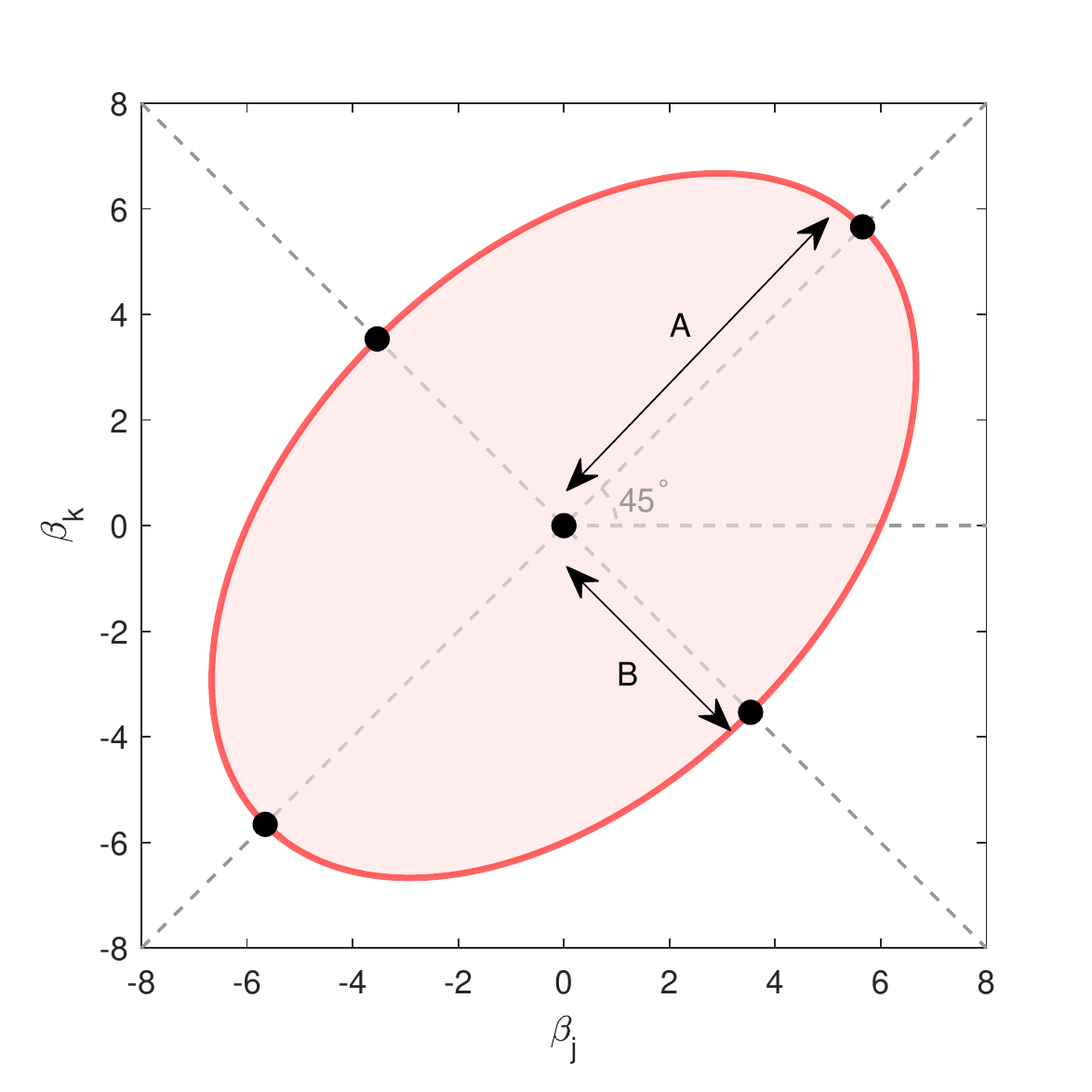}}
    \caption{(a) $f(\beta_2)$ in scenario 1 ($p=2$); (b) the projection of $f(\bb)=0$ in scenario 2 ($p\ge 3$).}
    \label{fig:ellipse}
\end{figure}

When $ \frac{\sigma_1^2\sigma_2^2}{\sigma_{12}^2}\ge2$ (i.e., $|\rho_{12}|\le\frac{\sqrt{2}}{2}$), the AC estimator has the possibility to be better than the CC as long as $\beta_2$ is not too far from~$0$. In Figure~\ref{fig:ellipse}(a), we plot function $f(\beta_2)$ and find that $f>0$ iff $\beta_2$ lies in the interval between two intersections (the pink interval). This interval is symmetric around $0$ and we denote its length as $C$:
\begin{align*}
    C 
    &= \sqrt{\frac{4\left(\sigma_1^2\sigma_2^2-2\sigma_{12}^2\right)\sigma^2}{\left(\frac{1+2\kappa}{1+\kappa}\sigma_{12}^2+\sigma_1^2\sigma_{2}^2\right)\sigma_2^2}}.
\end{align*}

\begin{table}[!t]
    \renewcommand{\arraystretch}{0.7}
    \newcolumntype{Y}{>{\centering\arraybackslash $}X<{$}}
    \caption{How $C$ changes with parameters}\label{tab:c}
    \begin{tabularx}{\textwidth}{YYYYY}
    \toprule
    \text{Parameter} &  \text{Segment  1} & C & \text{Segment 2} & C \\
    \midrule
    \rowcolor[gray]{0.9}\kappa &   (-1/2, +\infty) & \searrow &   &  \\
    \sigma_{12} &    (-\sqrt{\sigma_1^2\sigma^2_2/2}, 0) & \nearrow & (0, \sqrt{\sigma_1^2\sigma^2_2/2}) & \searrow \\
    (\rho_{12}) &    (-\sqrt{2}/2, 0) & \nearrow & (0, \sqrt{2}/2) & \searrow \\
      \rowcolor[gray]{0.9}\sigma_2^2 & (2\sigma_{12}^2/\sigma_1^2, M_0\ ^a) & \nearrow & (M_0, +\infty) & \searrow \\
    \sigma_1^2 &   (2\sigma_{12}^2/\sigma_2^2, +\infty) & \nearrow &   &  \\
    \rowcolor[gray]{0.9}\sigma^2 &  (0, +\infty) & \nearrow &   &  \\
    \bottomrule
\end{tabularx}\\
\spacingset{1}\footnotesize Note: $^a\ M_0 =  \left(2+\sqrt{8-2/(1+\kappa)}\right)\sigma_{12}^2/\sigma_1^2$\\
\end{table}

We list how $C$ changes with different parameters in Table~\ref{tab:c}. When the kurtosis parameter $\kappa$ increases, the interval length $C$ decreases, which means a heavy-tailed dataset favors the CC method. For the covariance structure, we find that a larger $\sigma_1^2$, $\sigma^2$ and a smaller $|\sigma_{12}|$ favors the AC estimator, but the effect of $\sigma_2^2$ is not monotone when fixing other parameters. In other words, increasing the variance of $X_1$ or the residual, and decreasing the correlation between $X_1$, $X_2$ make the AC estimator of $\beta_1$ has a smaller asymptotic variance. 
\\

\noindent\textbf{Scenario 2, $\bf p\ge3$:}

In this scenario, we assume that $X_j\,(j\ge2)$ are homoscedastic and has an exchangeable covariance structure. Their correlation with $X_1$ is exchangeable as well. Specifically, we assume that the variance of $X_1$ is $\sigma_1^2$; the variance of $X_j\,(j\ge2)$ is $\sigma_{2'}^2$; the covariance between $X_1$ and $X_j\,(j\ge2)$ is $\sigma_{12'}$; and the covariance between $X_j\,(j\ge2)$ and $X_k\,(k\ge2, k\ne j)$ is $\sigma_{2'3'}$. Then $f(\bb)$ is simplified as:
\begin{align*}
   f(\bb)=& -r_{11}^2(\sigma_{12'}^2+2\kappa \sigma_{12'}^2+\sigma_1^2\sigma_{2'}^2+\kappa \sigma_1^2\sigma_{2'}^2)\left(\frac{1}{q_{-1}}-\frac{1}{q_1}\right) \sum_{g=2}^p\beta_g^2\\
    &\qquad - 2r_{11}^2(\sigma^2_{12'}+2\kappa\sigma_{12'}^2+\sigma_1^2\sigma_{2'3'}+\kappa\sigma_1^2\sigma_{2'3'})\left( \frac{1}{q_{-1}}-\frac{1}{q_1} \right) \sum_{g=2}^p\sum_{h=g+1}^p \beta_g\beta_h \\
    & \qquad+ (1+\kappa)\left(\frac{1}{q_{-1}}-\frac{1}{q_1}\right)\left(2r_{11}-r_{11}^2\sigma_1^2\right)\sigma^2.
\end{align*}

We find that $f(\bb)$ is an elliptic paraboloid $\mathbb{R}^{p}$. When the constant term (that does not involve $\bb$) in $f(\bb)$ is negative, $f(\bb)$ is always negative (See Appendix F (online supplemental material) for proof). So we have the following proposition:
\begin{prop}
    (See Appendix E (online supplemental material) for proof) In missing pattern (a) with all assumptions above, a sufficient condition that the CC estimator of $\beta_1$ is asymptotically better than the AC is:
\begin{align*}
    \{(p-2)\sigma_{2'3'}+\sigma_{2'}^2\}\sigma_1^2<2(p-1)\sigma_{12'}^2.
\end{align*}
As $p\rightarrow \infty$, this condition becomes:
\begin{align*}
    \frac{\sigma_{2'3'}\sigma_1^2}{\sigma_{12'}^2}<2.
\end{align*}
\end{prop}

The condition $\sigma_{2'3'}\sigma_1^2<2\sigma_{12'}^2$ is equivalent to $\rho_{2'3'}<2\rho_{12'}^2$, where $\rho_{12'}$, $\rho_{2'3'}$ is the correlation between $X_1$, $X_j\,(j\ge2)$, and $X_j\,(j\ge 2)$, $X_k\,(k>1,k\ne j)$ respectively. 
This proposition shows that in a high dimensional dataset ($p$ is large) with missing pattern (a), if the correlation between $X_1$ and $X_j\,(j\ge 2)$ is too strong ($|\rho_{12'}|>\sqrt{\frac{|\rho_{2'3'}|}{2}}$), the AC estimator is always worse.

\begin{table}[!t]
    \renewcommand{\arraystretch}{0.7}
    \newcolumntype{Y}{>{\centering\arraybackslash $}X<{$}}
    \caption{How $A$ changes with parameters}\label{tab:a}
    \begin{tabularx}{\textwidth}{Y>{$}c<{$}YYYY}
    \toprule
    \text{Parameter} & \text{Condition} & \text{Segment 1} & A & \text{Segment 2} & A \\
    \midrule
    \rowcolor[gray]{0.9}p &   & (I_L, I_R)\ ^a & \searrow &   &  \\
    \kappa &   & (I_L, +\infty) & \searrow &   &  \\
    \rowcolor[gray]{0.9}\sigma_{12'} &   & (I_L, 0) & \nearrow & (0, I_R) & \searrow \\
    \multirow{3}[0]{*}{$\sigma_{2'3'}$} & I_L<M_1\ ^b<I_R & (I_L, 0) & \nearrow & (0, I_R)& \searrow \\
      & I_R<M_1 & (I_L, I_R) & \nearrow &   &  \\
      & I_L>M_1 & (I_L, I_R) & \searrow &   &  \\
      \rowcolor[gray]{0.9}& con\_1\ ^c>0 \text{ and }I_L<M_2\ ^b & (I_L, M_2) & \nearrow & (M_2, +\infty) & \searrow \\
    \rowcolor[gray]{0.9}\multirow{-2}[0]{*}{$\sigma_{2'}^2$}   & con\_1<0 \text{ or }I_L>M_2 & (I_L, +\infty) & \searrow &   &  \\
    \sigma_1^2 &   & (I_L, +\infty) & \nearrow &   &  \\
    \rowcolor[gray]{0.9}\sigma^2 &   & (0, +\infty) & \nearrow &   &  \\
    \bottomrule
\end{tabularx}\\
\spacingset{1}\footnotesize Note: $^a\ I_L, I_R$ are the minimum/maximum value for this parameter to take (See Appendix G (online supplemental material))\\
\phantom{Note: }$^b$ The expressions of  $M_1, M_2$ are in Appendix G (online supplemental material)\\
\phantom{Note: }$^c\ con\_1=(2+2p-2/(1+\kappa))\sigma_{12'}^2+(3-p)\sigma_{1}^2\sigma_{2'3'}$

    \caption{How $B$ changes with parameters}\label{tab:b}
    \begin{tabularx}{\textwidth}{Y>{$}c<{$}YYYY}
    \toprule
    \text{Parameter} & \text{Condition} & \text{Segment 1} & B & \text{Segment 2} & B \\
    \midrule
    \rowcolor[gray]{0.9}p &   & (I_L, I_R)\ ^a & \searrow &   &  \\
    \kappa &   & (I_L, +\infty) &  \rightarrow &   &  \\
    \rowcolor[gray]{0.9}\sigma_{12'} &   & (I_L, 0) & \nearrow & (0, I_R) & \searrow \\
    \multirow{2}[0]{*}{$\sigma_{2'3'}$} & 2\sigma_{12'}^2>\sigma_{1}^2\sigma_{2'}^2 \text{ and } I_R>M_3\ ^b & (I_L, M_3) & \nearrow & (M_3, I_R) & \searrow \\
      & 2\sigma_{12'}^2<\sigma_{1}^2\sigma_{2'}^2 \text{ or } I_R<M_3 & (I_L, I_R) & \nearrow &   &  \\
    \rowcolor[gray]{0.9} & 2\sigma_{12'}^2>\sigma_{1}^2\sigma_{2'3'} \text{ and } I_L<M_4\ ^b & (I_L, M_4) & \nearrow & (M_4, +\infty) & \searrow \\
    \rowcolor[gray]{0.9}\multirow{-2}[0]{*}{$\sigma_{2'}^2$}  & 2\sigma_{12'}^2<\sigma_{1}^2\sigma_{2'3'} \text{ or } I_L>M_4 & (I_L, +\infty) & \searrow &   &  \\
    \sigma_1^2 &   & (I_L, +\infty) & \nearrow &   &  \\
    \rowcolor[gray]{0.9}\sigma^2 &   & (0, +\infty) & \nearrow &   &  \\    
    \bottomrule
\end{tabularx}\\
\spacingset{1}\footnotesize Note: $^a\ I_L, I_R$ are the minimum/maximum value for this parameter to take (See Appendix G (online supplemental material))\\
\phantom{Note: }$^b$ The expressions of  $M_3, M_4$ are in Appendix G (online supplemental material)\\
\end{table}
In Figure~\ref{fig:ellipse}(b), we plot this ellipse whose center is at the origin and the major axis is rotated $45^\circ$ around the origin. When point $(\beta_j,\beta_k)$ lies in the ellipse (the pink region), then the AC estimator is better than the CC. Let $A$ and $B$ denote the length of the semi-major and semi-minor axes:
\begin{align*}
  A
  &=\sqrt{\frac{\left(-2(p-1)\sigma_{12'}^2+(p-2)\sigma_1^2\sigma_{2'3'}+\sigma_1^2\sigma_{2'}^2\right)\sigma^2}{\left((p-2)\sigma_{2'3'}+\sigma_{2'}^2\right)\left(\frac{2+4\kappa}{1+\kappa}\sigma_{12'}^2+\sigma_1^2\sigma_{2'}^2+\sigma_1^2\sigma_{2'3'}\right)}},\\
  B
  &=\sqrt{\frac{\left(-2(p-1)\sigma_{12'}^2+(p-2)\sigma_1^2\sigma_{2'3'}+\sigma_1^2\sigma_{2'}^2\right)\sigma^2}{\left((p-2)\sigma_{2'3'}+\sigma_{2'}^2\right)\left(\sigma_1^2\sigma_{2'}^2- \sigma_1^2\sigma_{2'3'}\right)}}.
\end{align*}

Similar to scenario 1, when $((p-2)\sigma_{2'3'}+\sigma_{2'}^2)\sigma_1^2\ge2(p-1)\sigma_{12'}^2$, the AC method has potential to be better than the CC. To be more specific, if setting $f(\bb)=0$, we get an ellipsoid in $\mathbb{R}^{p-1}$ space. This ellipsoid is symmetric around the origin and its projection onto any $(\beta_j,\beta_k)$-plane has the same shape and size. The projection curve on the $(\beta_j,\beta_k)$-plane is an ellipse and described by the following expression: 
\begin{align*}
   & \left(\frac{1+2\kappa}{1+\kappa}\sigma_{12'}^2+\sigma_1^2\sigma_{2'}^2\right)(\beta_j^2+\beta_k^2)
   + 2\left(\frac{1+2\kappa}{1+\kappa}\sigma^2_{12'}+\sigma_1^2\sigma_{2'3'}\right) \beta_j\beta_k 
    =\left(\frac{2}{r_{11}}-\sigma_1^2\right)\sigma^2 .
\end{align*}

We list how $A$, $B$ change with different parameters in Table~\ref{tab:a} and \ref{tab:b}. In particular, when the number of predictors $p$ increases, both axes get shorter, resulting in a smaller ellipse that favors the CC method. 
Larger kurtosis parameter $\kappa$ also shrinks the ellipse, which means a heavy-tailed dataset impairs the performance of AC.  
In addition, we find that a larger $\sigma_1^2$, $\sigma^2$ and a smaller $|\sigma_{12'}|$ favor AC estimator. The effect of $\sigma_{2'}^2$, $\sigma_{2'3'}$ is not monotone. We conclude that a lower correlation between $X_1$ and other predictors, a larger variance of $X_1$ or the residual benefit the AC estimator.\\

\subsubsection{Missing Pattern (b)}\label{sec:inv_mono}
This missing pattern is shown in Figure~\ref{fig:missing}(b). Let $q_1$ denote the observed proportion of $X_1$; $q_{-1}$ be the observed proportion of $X_j (j\ge 2)$. In addition, we assume available samples in $X_1$ is a subset of $X_2$ to $X_p$ ($q_1< q_{-1}$). A special case is that  only variable $X_1$ has missing values ($q_{-1}=1$) which is called \textit{univariate missing}. With expressions~\eqref{eqn:singleac}, \eqref{eqn:singlecc}, we obtain the asymptotic variance of $\hat\beta_1$ of two methods and 
the difference $V_{D,\hat\beta_1}$ is as follows:
\begin{align*}
    n\cdot V_{D,\hat\beta_1} 
    & = \left(\frac{1}{q_1}-\frac{1}{q_{-1}}\right)\left\{n(r_{11}-r_{11}^2\sigma_1^2)\sigma^2-c_1\beta_1^2\right\},
\end{align*}
where 
\begin{align*}
    c_1&=\sum_{\substack{j=2}}^p r_{1j}^2\left( \sigma_{1j}^2+\sigma_1^2 \sigma_j^2\right)
    +\kappa r_{1j}^2\sigma_1^2 \sigma_j^2
    +2\kappa r_{1j}^2\sigma_{1j}^2\\
    &\qquad \quad+\sum_{\substack{j=2}}^p \sum_{\substack{k=j+1}}^p 2r_{1j}r_{1k}\left( \sigma_{1j}\sigma_{1k}+\sigma^2_1\sigma_{jk} \right) 
    +2\kappa r_{1j}r_{1k}\sigma^2_1\sigma_{jk}
    + 4\kappa r_{1j}r_{1k}\sigma_{1j}\sigma_{1k}.
\end{align*}
The asymptotic variance of $\hat\beta_{1}$ in the CC method is always equal or smaller than the AC method (See Appendix H (online supplemental material) for proof). The quantity $\left(\frac{1}{q_{1}}-\frac{1}{q_{-1}}\right)$ determines the difference of performance between two methods. 

\begin{prop}
In missing pattern (b), the CC estimator of $\beta_1$ is asymptotically equal or better than the AC estimator.
\end{prop}
This proposition implies that using extra data from $X_2$ to $X_p$ does not improve the estimation of $\beta_1$ asymptotically. The special case is that when $X_1$ is independent with other predictors, then both $(r_{11}-r_{11}^2\sigma_1^2)=0$ and $c_1=0$ and thus we have the following proportion:
\begin{prop}
    (See Appendix H (online supplemental material) for proof) In missing pattern (b), the AC and the CC have the same asymptotic performance if and only if $X_1$ is independent with other predictors.
\end{prop}

\subsection{Summary}\label{sec:summary}

\begin{table}[!t]
    \renewcommand{\arraystretch}{0.7}
\newcolumntype{Y}{>{\centering\arraybackslash}X}
\caption{Summary table of the comparison in different scenarios. }\label{tab:summary}
\begin{tabularx}{\textwidth}{cccYY}
    \toprule
Missing Pattern    & $p$&  Condition & AC & CC \\
    \midrule
    \multirow{7}[0]{*}{pattern (a)} &   & all predictors are independent, $\sum_{g=2}^p\sigma_g^2\beta_g^2<\sigma^2$ & $\checkmark$ &  \\
      & $2$ & $|\rho_{12}|>\sqrt{2}/2$ &   & $\checkmark$ \\
      & $2$ & large $\kappa$, $|\rho_{12}|$ &   & $\bigstar$ \\
      & $2$ & large $\sigma_1^2$, $\sigma^2$; small $|\beta_2|$ & $\bigstar$ &  \\
      & $\ge 3$ & special covariance $^a$; $\rho_{12'}^2>\rho_{2'3'}/2$ $^b$ (large $p$) &   & $\checkmark$ \\
      & $\ge 3$ & special covariance; large $p$, $\kappa$, $|\sigma_{12'}|$ &   & $\bigstar$ \\
      & $\ge 3$ & special covariance; large $\sigma_1^2$, $\sigma^2$; small $|\beta_g|$ ($g\ge 2$) & $\bigstar$ &  \\
      \hdashline
    \multirow{2}[0]{*}{pattern (b)} &   & $X_1$ is independent with other predictors & \multicolumn{2}{c}{same}  \\
      &   & $X_1$ is not independent with other predictors &   & $\checkmark$ \\ 
      \bottomrule  
\end{tabularx}\\
\spacingset{1}\footnotesize Note: $^a$ $X_j (j\ge2)$ are homoscedastic and has an exchangeable covariance structure. Their correlation with $X_1$ is exchangeable as well.\\
\phantom{Note: }$^b$ This condition becomes $\{(p-2)\sigma_{2'3'}+\sigma_{2'}^2\}\sigma_1^2<2(p-1)\sigma_{12'}^2$ when $p$ is not large enough.\\
\phantom{Note: }$\checkmark$ represents the better estimator in this condition; $\bigstar$ represents this condition flavors the method, but it is not guaranteed to be better. 
\end{table}
The main results of Subsection~\ref{sec:comparison_single} are listed in Table~\ref{tab:summary}. In missing pattern (a) that available samples in other predictors is a subset of $X_1$, the CC estimator of $\beta_1$ outperforms the AC estimator when the correlation between $X_1$ and others are large (i.e., $|\rho_{12}|>\sqrt{2}/2$ for two-dimensional predictors;  $\rho_{12'}^2>\rho_{2'3'}/2$ for very high dimensional predictors with special covariance structure). A larger variance of $X_1$ and residual, and true coefficients $\beta_g$ ($g\ge 0$) that is closer to $0$ increase the relative performance of the AC; while higher dimension of predictors, heavier distribution tails, and larger correlation between $X_1$ and other predictors flavor the performance the CC. In missing pattern (b) that available samples in $X_1$ is a subset of other predictors, the CC estimator of $\beta_1$ is better, except for the scenario that when $X_1$ is independent with other predictors, both methods have the same asymptotic performance.

\section{SIMULATION STUDY}\label{sec:simulation}
In the introduction part, we mentioned several simulation studies that tried to evaluate the AC method. In this section, we take \citet{kim1977treatment}'s paper as an example to illustrate the performance of the AC method comparing with the CC in more details with the help of our theoretical results from the last section. 

The simulation studies are based on the correlation matrix on page 196 in \citet{blau1967american}'s book. All results from \citet{kim1977treatment} showed that the AC method is superior to the CC. For example, there is a regression analysis of education status. Response variable $U$ is education status and predictors are $V$ (father's education) and $X$ (father's occupational status):
\begin{align*}
    U = 0.310V+0.279X+\varepsilon.
\end{align*}

The variance of two predictors and the residual is $1$ ($\sigma_V^2=\sigma_X^2=\sigma_\varepsilon^2=1$). The covariance between $V$, $X$ is $0.516$ ($\sigma_{VX}=0.516$). So we obtain the covariance matrix of random vector $(V,X)^T$ and $(V,X,U)^T$ as follows:
\begin{align*}
    \sb_{VX}=\begin{pmatrix}
        1 & 0.516\\
        0.516 & 1
    \end{pmatrix},\qquad \sb_{VXU}=\begin{pmatrix}
        1 & 0.516& 0.454\\
        0.516 & 1 & 0.439\\
        0.454 & 0.439 & 1.263
    \end{pmatrix}.
\end{align*}

\subsection{Finite-Sample Performance}
In the first part of the simulation study, we set up five settings to examine the asymptotic property of our theoretical results. 

\begin{itemize}[parsep=0pt,itemsep=0pt,topsep=0pt,labelindent=\parindent,leftmargin=*]
	\item \textbf{Setting (1)}. $(V, X, U)^T\sim N(\boldsymbol{0},\sb_{VXU})$. This setting assumes predictors and errors are normally distributed.
    \item \textbf{Setting (2)}. $(V, X, U)^T\sim t_5(\boldsymbol{0},\frac{3}{5}\sb_{VXU})$. In this setting, the covariance matrix of $(V, X, U)$ is $\sb_{VXU}$. The response and predictors follow a multivariate t distribution with degree of freedom of $5$, which meets the elliptic distribution assumption. 
	\item \textbf{Setting (3)}. $(V, X)^T\sim$ Multivariate Bernoulli with covariance matrix $\sb_{VX}/6$. The response $U=0.310V+0.279X+\varepsilon$ where $\varepsilon\sim N(0,1)$. This is a typical case where predictors are categorical variables but the error term is normally distribution. This setting violates the elliptic distribution assumption.
	\item \textbf{Setting (4)}. $(V, X)^T\sim$ Multivariate Poisson with covariance matrix $\sb_{VX}$. The response $U=0.310V+0.279X+\varepsilon$ where $\varepsilon\sim N(0,1)$. Another setting that violates the elliptical distribution assumption. The setting is similar to Setting (3), but the predictors follow a multivariate Poisson distribution that has a larger kurtosis than Bernoulli. 
	\item \textbf{Setting (5)}. $(V, X, U)^T\sim$ Multivariate Poisson with covariance matrix $\sb_{VXU}$. Comparing with Setting (4), the error does not follow a normal distribution. This also violates the elliptical distribution assumption.
\end{itemize}
In each setting, each predictor has $10\%$ of random missing cases. Without loss of generality, we focus on the variance of coefficient estimator for predictor~$V$ (i.e., $\hat\beta_V$). We calculate the variance of $\hat\beta_V$ with $10,000$ estimated $\hat\beta_V$ in the AC method and repeat the simulations 100 times to obtain the standard deviation. The sample size varies from $50$ to $250$. The theoretical kurtosis parameter $\kappa$ is estimated from the samples using the second approach in Appendix D (online supplemental material). 

In Figure~\ref{fig:asy}, we plot the theoretical results in solid lines and simulated results in dashed lines. In the first two settings where the elliptical distribution assumption holds, the theoretical variance converges to the simulated variance quickly and can be used to represent the true variance accurately when sample size is larger than $150$. 
For the rest of settings, the theoretical result slightly overestimates the variance in Setting (4) and performs well in Setting (3) and (5). The main reason is that the kurtosis of response $U$ is quite different from that of $V$, $X$ in setting (4). The kurtosis parameter $\kappa$ is estimated as one third of the mean excess kurtosis of each variable and thus its value is not accurate in setting (4), which results in a worse convergence property. Overall, we conclude that  it is safe to utilize the theoretical asymptotic variance to analysis the true estimator variance when the elliptical distribution assumption holds and sample size is not too small. When the assumption severely violates, the derived asymptotic variance may be inaccurate in some cases, especially when the kurtosis of each variable varies much.

\begin{figure}[!ht]
\centering
    \includegraphics[width=.79\textwidth]{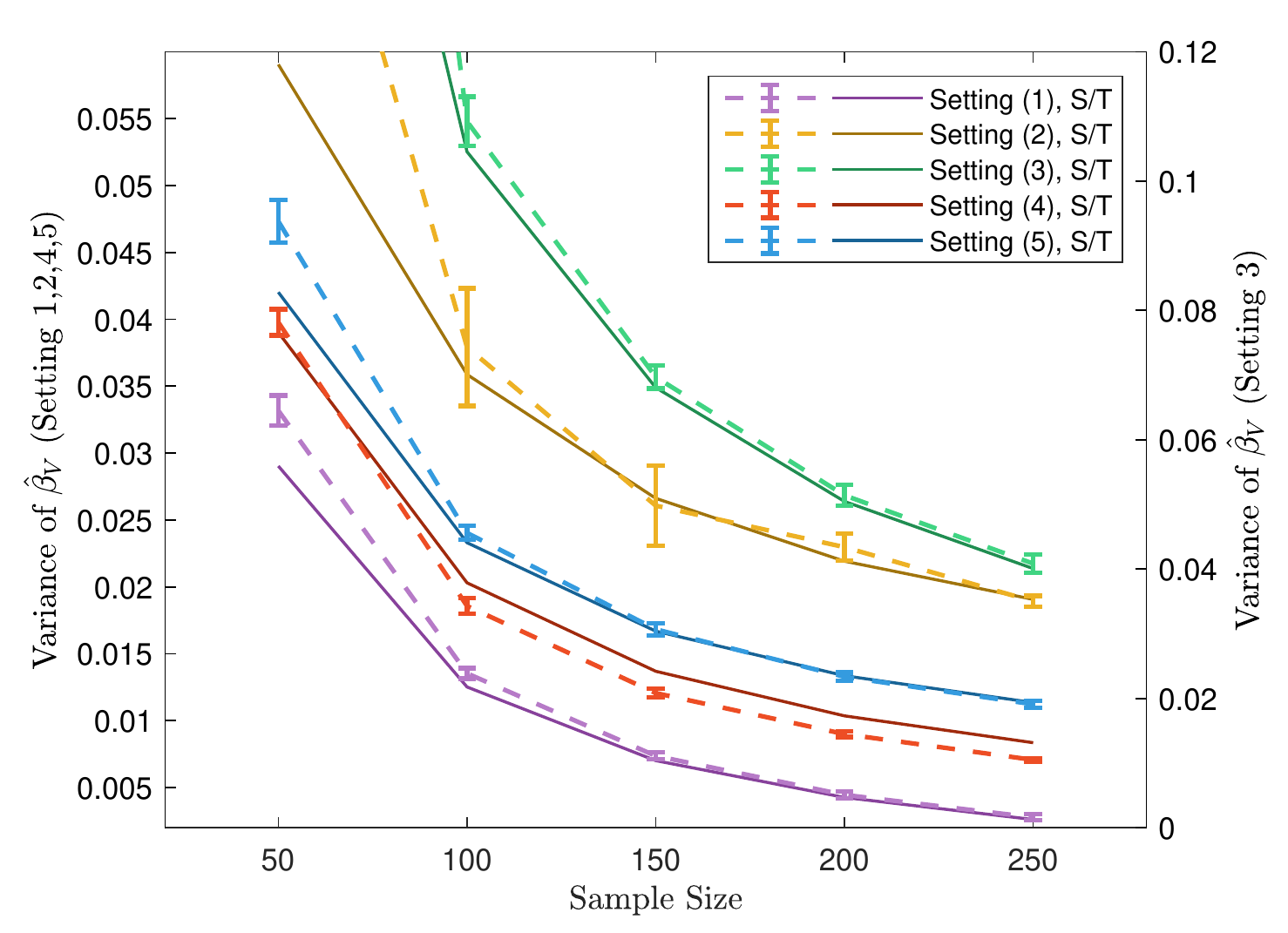}
    \caption{Asymptotic performance of the theoretical results. ``S/T'' stand for simulated/theoretical result respectively.}\label{fig:asy}
\end{figure}

\subsection{Comparison of AC and CC}
In the second part of the simulation study, we still focus on the variance of $\hat\beta_V$. Here, we try to explore how the performance of the AC changes with different model parameters when comparing with the CC estimator. The response and predictors are simulated from a multivariate-normal distribution as in Setting (1) with $1000$ ($n=1000$) samples.

\begin{table}[!b]
    \renewcommand{\arraystretch}{0.7}
    \newcolumntype{Y}{>{\centering\arraybackslash $}X<{$}}
    \caption{Different missing patterns}\label{tab:missing_p}
    \begin{tabularx}{\textwidth}{>{$}c<{$}>{$}c<{$}>{$}c<{$}YYYY}
        \toprule
        q_V & q_X & q_{VX} & \var(\hat\beta_{V,CC}^{\text{(s)}})^a & \var(\hat\beta_{V,CC}^{\text{(t)}})^a  & \var(\hat\beta_{V,AC}^{\text{(s)}})^a & \var(\hat\beta_{V,AC}^{\text{(t)}})^a\\
        \midrule
        0.9 & 0.9 & 0.81 & 1.6820 & 1.6826  & \textbf{1.6312} & \textbf{1.6516} \\
        0.9 & 1 & 0.9 & \textbf{1.4952} & \textbf{1.5143}  & 1.5543 & 1.5759 \\
        0.8 & 1 & 0.8 & \textbf{1.7037} & \textbf{1.7036}  & 1.8409 & 1.8423 \\
        1 & 0.9 & 0.9 & 1.4952 & 1.5143  & \textbf{1.4313} & \textbf{1.4382} \\
        1 & 0.8 & 0.8 & 1.7037 & 1.7036  & \textbf{1.5294} & \textbf{1.5323} \\
        \bottomrule       
    \end{tabularx}\\
    \spacingset{1}\footnotesize Note: $^a$ (s) means simulation results; (t) means theoretical results using expressions~\eqref{eqn:singleac}, \eqref{eqn:singlecc}. All the variances are in the order of magnitude of $-3$.\\
    \phantom{Note: }The smaller variances in each setting are in bold face.
\end{table}

In Table~\ref{tab:missing_p} row $1$, we reproduce the result of the \citet{kim1977treatment}'s setting where they deleted $10\%$ of the cases from both predictors randomly and obtain their finding that the AC methods has the smaller variance. Then we explore the simpler settings that only one predictor has missing values. In row 2--3, only $V$ has missings and as we discussed in Subsection~\ref{sec:inv_mono},  the AC method will not improve the efficiency of $\hat\beta_V$ by using extra data from $X$. Therefore, the CC method is always better in this missing pattern. The last two rows are the settings where $X$ has missings. The AC method even has larger advantages than \citet{kim1977treatment}'s setting. In addition, we observe that no matter which method is better, the performance difference gets larger when the missing proportion increases, which is consistent with finding that the inverse of the observation proportion serves as a scalar in $V_{D}$ in Subsection~\ref{sec:mono}, Subsection~\ref{sec:inv_mono}. 

\begin{figure}[!b]
    \centering
    \parbox{0.49\textwidth}{\textbf{(a)}\\
        \includegraphics[width=.5\textwidth]{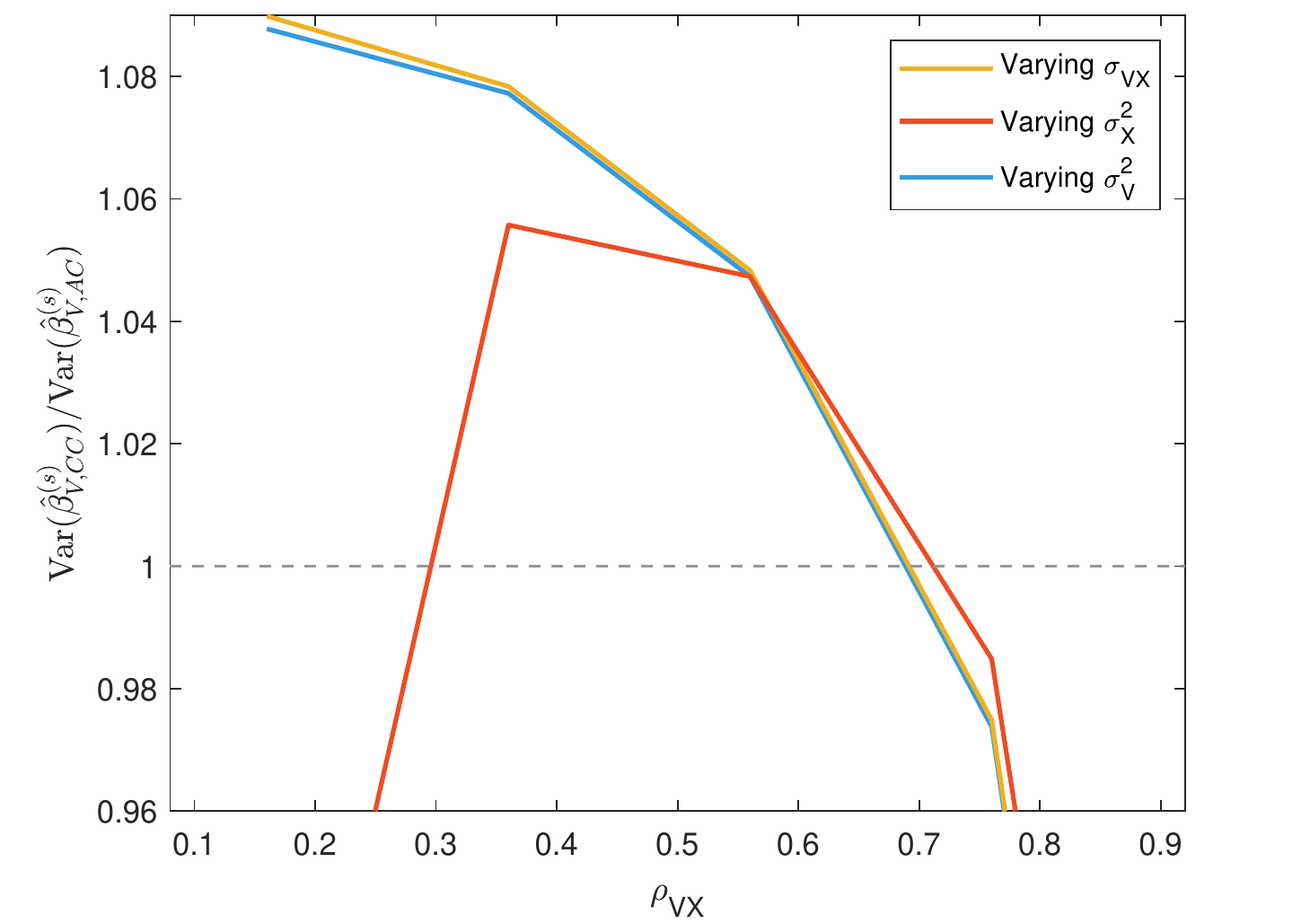}
    }
    \parbox{0.49\textwidth}{\textbf{(b)}\\
        \includegraphics[width=.5\textwidth]{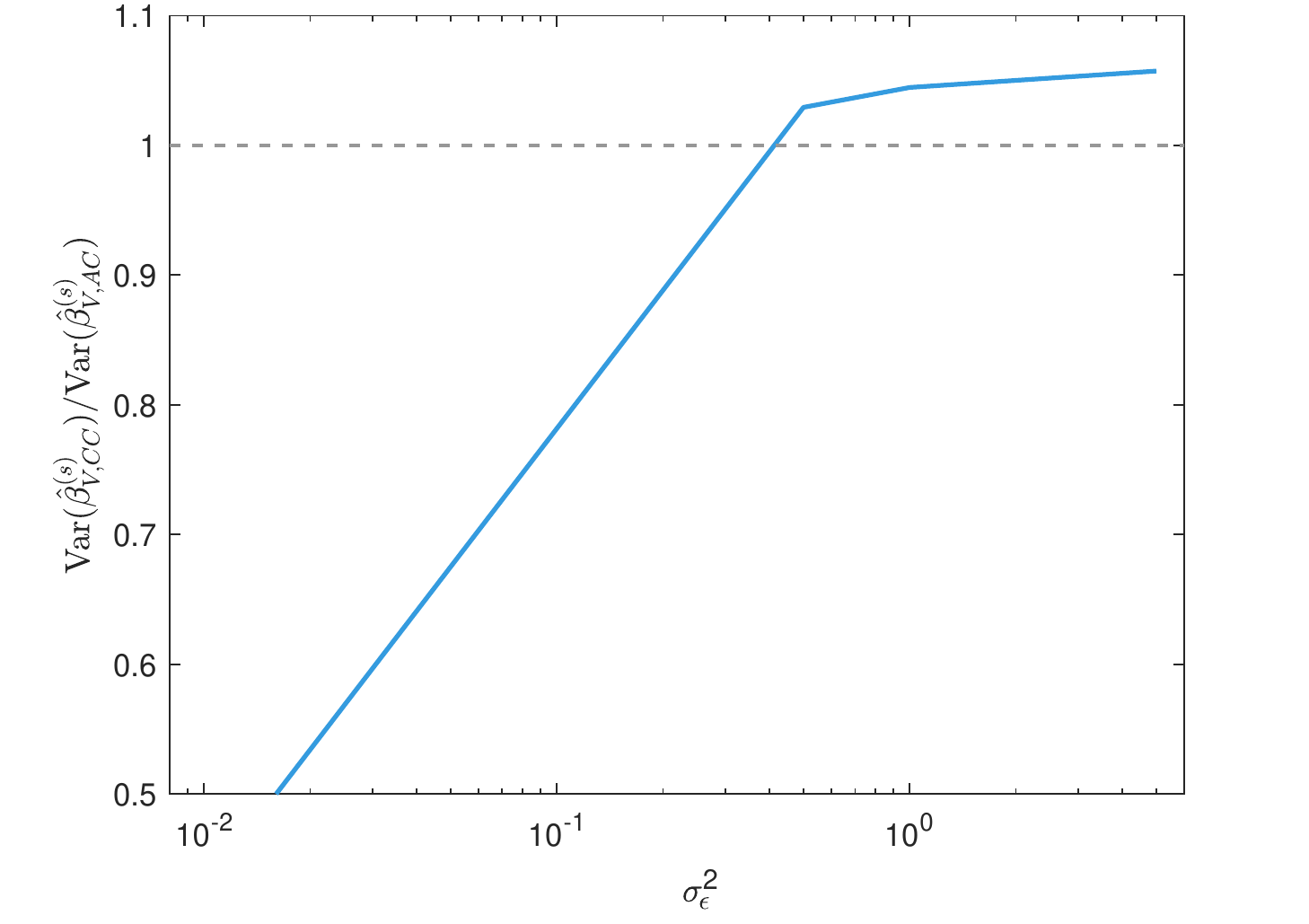}
    }
    \parbox{0.49\textwidth}{\textbf{(c)}\\
    \includegraphics[width=.5\textwidth]{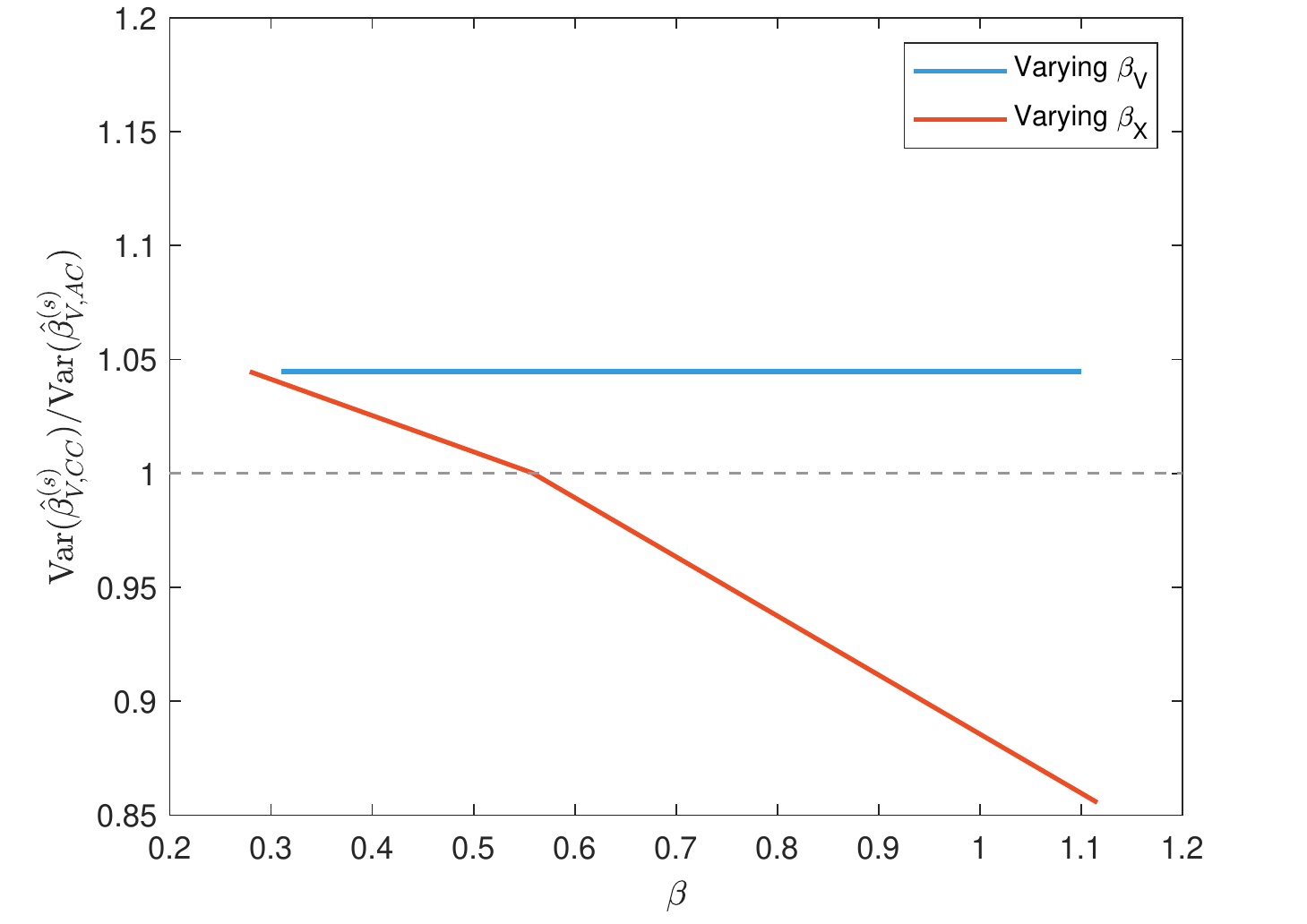}
}
\caption{The relative efficiency of $\hat \beta_V$ between CC, AC. Single parameter changes each time. (a) Different  correlations of $V$, $X$. (b) Different residual variances. (c) Different coefficients.}\label{fig:varying}
\end{figure}

We further investigate the influence of different parameters on the performance of different methods. We focus on the setting with complete $V$ and incomplete $X$ with $10\%$ missing (i.e., row 4 in Table~\ref{tab:missing_p}). The results are presented in Appendix Table 1 to 3 (online supplemental material) and the relative efficiency of $\hat \beta_V$ between CC, AC is shown in Figure~\ref{fig:varying}. In Appendix Table~1 (online supplemental material), we change the correlation between $V$, $X$ in different ways. We fix the variance of $V$ and $X$ in the first part of the table, and it turns out that the AC method outperforms the CC more when the covariance decreases (yellow line in Figure~\ref{fig:varying}(a)). Some articles misinterpret this result and claim that it is better to use the AC method when the correlation between predictors is small \citep{glasser1964linear,kim1977treatment}. The counterexample is in the second part of the table, where we decrease the correlation by increasing the variance of $X$. As we point out in Subsection~\ref{sec:mono}, the effect of $\sigma_X^2$ is not monotone, so that we see the CC method beats the AC in both low and high correlation situations (red line in Figure~\ref{fig:varying}(a)). Lastly, we decrease the correlation by increasing the variance of $V$ in the third part. As expected, the effect of $\sigma_V^2$ is monotone. A larger variance of $V$ flavors the AC method (blue line in Figure~\ref{fig:varying}(a)). 

The effect of $\sigma_\varepsilon^2$ is simple. As shown in Appendix Table~2 (online supplemental material) and Figure~\ref{fig:varying}(b), a larger variance of the residual makes the AC methods more advantageous. 

The most severe problem of the AC method is showed in Appendix Table~3 (online supplemental material). We use different true coefficients to compare two methods. From the theoretical results (expressions~\eqref{eqn:singleac}, \eqref{eqn:singlecc}), we know that the asymptotic variance of coefficients are not related with the true coefficients in the CC methods, but that will change in the AC method. 
In this setting, the variance of estimated $\hat\beta_V$ in the AC method increases with true $\beta_X$ increases and thus relative efficiency decreases (red line in Figure~\ref{fig:varying}(c)). 
Therefore, the AC method will be inferior when the effect size of other predictors are large. Especially when there are several predictors, the AC method is worse as long as any of the other coefficients is large.

\section{DISCUSSION}
Since both the AC estimator and the CC estimator are asymptotically unbiased under the MCAR assumption, the comparison of their asymptotic variance helps us determine which one is better. 
The missing pattern, data covariance structure and true coefficients together influence the performance comparison and their effects on the estimator efficiency under two special missing patterns are summarized in Subsection~\ref{sec:summary}. Generally speaking, the AC estimator has advantages when the predictors are not strongly correlated, especially for the variable of interest. A heavier tailed data distribution and higher predictor dimension flavor the CC estimator. 

We find that the comparison outcome is quite complicated and hope to have some more intuitive explanations. \citet{tarpey2014paradoxical} pointed out that the least-squares estimator $\hat\bb=\Sx^{-1}\Sxy$ enjoys special superiority because it is a ratio estimator that exploits the dependence between $\Sx$ and $\Sxy$, so that even the estimator $\tilde\bb=\sx^{-1}\Sxy$ that uses the true $\sx$ to replace $\Sx$ is inferior to the least-squares estimate $\hat\bb$. Therefore, though the AC estimator utilizes more data than the CC method, it is not always better because it may impair the benefits from utilizing dependence between $\Sx$ and $\Sxy$. This argument provides a potential perspective on the intuitive explanation and needs to be formalized in the further work. 

Another possible research direction is to extend the work to more scenarios. We can further relax the elliptical assumption, include more general missing patterns, or compare several coefficients at the same time. These works will help us to gain insight into the whole picture and lead to more elegant solutions.

Besides, the AC estimator has many improved versions \citep{yu2020optimal, lorenzo2020not} that possibly have better performance than the original one. Future researches can evaluate these variants and further optimize the method. 

\section*{ACKNOWLEDGEMENTS}
We thank Dr. Rod Little for helpful comments. 
Gen Li's work was partially supported by National Institutes of Health [grant number R01HG010731]; Kun Chen's work is partially supported by National Science Foundation, Alexandria, VA [grant number IIS-1718798]. 

\section*{SUPPLEMENTARY MATERIALS}
\begin{description}

    \item[Proof and Simulation:] It mainly contains the proof of some propositions in Section~\ref{sec:comparison}, detailed simulation results in Section~\ref{sec:simulation}. (.pdf file)

    \end{description}
\section*{REFERENCE}
Reference is at the end of this document.

\newtheorem{con}{Conclusion}
\allowdisplaybreaks

\title{\bf Supplementary Materials for ``The More Data, the Better? Demystifying Deletion-Based Methods in Linear Regression with Missing Data" \\by}

\author{Tianchen Xu, Kun Chen, Gen Li }
\date{}
\maketitle

\setcounter{section}{0}
\section{Appendix A}
Define $\Sb^{*}$:
\begin{align*}
    \Sb^{*}=\vec\left[\begin{pmatrix}
        s_{{1}{1}} & s_{{1}{2}} &s_{{1}{3}} &\cdots &\cdots & s_{{1}{(p+1)}}\\
        Na & s_{{2}{2}} &s_{{2}{3}} &\cdots &\cdots & s_{{2}{(p+1)}}\\
        Na & Na &s_{{3}{3}} &\cdots &\cdots & s_{{3}{(p+1)}}\\
        &  &   &\ddots&  &\vdots \\
        Na & Na &Na  &Na&s_{{p}{(p)}} & s_{{p}{(p+1)}}\\ 
    \end{pmatrix}\right].
\end{align*}
Suppose we have a function $f: \mathbb{R}^{\frac{p^2+3p}{2}}\mapsto \mathbb{R}^{p}$:
\begin{align*}
    \hat \bb = f(\Sb^{*}).
\end{align*}
Then by the delta method:
\begin{align*}
    \sqrt{n}(\hat \bb - \bb)\god N_p(\boldsymbol{0}, \Deltab \Phib \Deltab^T)
\end{align*}
where 
\begin{align*}
    \Phib &= \left[\e(\Zb_j-\mu_j)(\Zb_k-\mu_k)(\Zb_m-\mu_m)(\Zb_n-\mu_n)-\sigma_{jk}\sigma_{mn}\right]\in \mathbb{R}^{\frac{p^2+3p}{2}\times \frac{p^2+3p}{2}},\\
    \Deltab &= \left[\frac{-\sb^{-1}_{x(.j)}\bb_k-\sb^{-1}_{x(.k)}\bb_j}{1+\mathbb{I}(j=k)}, \; \sx^{-1}\right]\in \mathbb{R}^{p\times \frac{p^2+3p}{2}}\\
    &= \sx^{-1}\begin{pmatrix}
        -\operatorname{diag}(\bb) \mathcal{P}_\beta,\boldsymbol{I}_p
    \end{pmatrix},
\end{align*}
with 
\setcounter{MaxMatrixCols}{30}
\begin{align*}
    \mathcal{P}_\beta = \begin{pmatrix}
        1   & 1   & 1   & 1   & \cdots & 1   &     &     &     &     &     &     &     &     &     &     &     &     &  \\
        & 1   &     &     &     &     & 1   & 1   & \cdots & 1   &     &     &     &     &     &     &     &     &  \\
        &     & 1   &     &     &     &     & 1   &     &     & 1   & 1   & \cdots & 1   &     &     &     &     &  \\
        &     &     & 1   &     &     &     &     & \ddots &     &     & 1   &     &     & 1   & \cdots & 1   &     &  \\
        &     &     &     & \ddots &     &     &     &     &     &     &     & \ddots &     &     &  \ddots   &     &     &  \\
        &     &     &     &     & 1   &     &     &     & 1   &     &     &     & 1   &     &     & 1   & \cdots & 1 \\  
    \end{pmatrix}_{p\times \frac{p(p+1)}{2}}.
\end{align*}

\section{Appendix B}
We prove this proposition by contradiction. Suppose there exists a matrix $\Qb$ such that $V_D$ is always positive definite for any $\Deltab$, $\Phib$.

Since $\Phib$ is the asymptotic variance of the limiting distribution of $\hat \bb$, then $\Phib$ is positive definite or positive semidefinite. By lemma~1.1, $(1-\Qb \tilde q)$ is not positive definite (might be semidefinite or indefinite). Therefore, $\Phib\circ (1-\Qb \tilde q)$ is not positive definite for some $\Phib_0$ by theorem 3.6 to theorem 3.8 in \citet{styan1973hadamard}'s paper. We know $\Deltab^T$ is full column rank since it involves $\sx^{-1}$ as a partition. By lemma~2, there exists $\Deltab_0$ such that $\Deltab_0 \left\{\Phib_0\circ (1-\Qb \tilde q)\right\}\Deltab_0^T $ is not positive definite, which is contradictory to our assumption that $\Deltab \left\{\Phib\circ (1-\Qb \tilde q)\right\}\Deltab^T$ is always positive definite for any $\Deltab$, $\Phib$.

The statement that $V_D$ is not negative definite for any $\Qb$ can be proved in a similar manner, except that we use lemma 1.2 instead of 1.1.\\



\noindent\textbf{Lemma 1.1:} Matrix $(1-\Qb \tilde q)$ is not positive definite for any $\Qb$ and $\tilde q$. \\
\noindent \textit{Proof}: Following the arrangement of $\Sb$ in Appendix~A, we can write down the second order leading principal submatrix $\boldsymbol{A_{(2)}}$ of $(1-\Qb \tilde{q})$:
\begin{align*}
    \boldsymbol{A_{(2)}} = \begin{dcases}
        \begin{pmatrix}
            1-\tilde{q}/q_1 & 1-\tilde{q}/q_1\\
            1-\tilde{q}/q_1 & 1-\tilde{q}/q_1 
        \end{pmatrix}, & p=1\\
        \begin{pmatrix}
            1-\tilde{q}/q_1 & 1-\tilde{q}/q_1\\
            1-\tilde{q}/q_1 & 1-\tilde{q}/q_{12} 
        \end{pmatrix}, & p\ge 2.\\
    \end{dcases}
\end{align*}
So the second order leading principal minor:
\begin{align*}
    |\boldsymbol{A_{(2)}}| = \begin{dcases}
      0, & p=1\\
        -\frac{(q_1-\tilde q)(q_1-q_{12})\tilde q}{q_1^2 q_{12}}, & p\ge 2.\\
    \end{dcases}
\end{align*}
is not positive, which implies $(1-\Qb \tilde q)$ cannot be a positive definite matrix. \\

\noindent\textbf{Lemma 1.2:} Matrix $(\Qb \tilde q-1)$ is not positive definite for any $\Qb$ and $\tilde q$. \\
\noindent \textit{Proof}: Following the arrangement of $\Sb$ in Appendix~A, the second order leading principal submatrix of $(\Qb \tilde{q}-1)$ is same as that of $(1-\Qb \tilde{q})$. Following the similar argument in lemma~1.1, we conclude $(\Qb \tilde q-1)$ cannot be a positive definite matrix. \\

\noindent\textbf{Lemma 2:} If symmetric matrix $\boldsymbol{A}$ is not positive definite then exists a full column rank matrix $\boldsymbol{B}^T$ such that $\boldsymbol{B}\boldsymbol{A}\boldsymbol{B}^T$ is not positive definite.\\
\noindent \textit{Proof}: This is just a simple corollary from the theorem 4 \citep{giorgi2017various} that symmetric matrix $\boldsymbol{A}$ is positive definite iff $\boldsymbol{B}^T\boldsymbol{A}\boldsymbol{B}$ is positive definite for any $n\times m$ matrix $\boldsymbol{B}$ with $\operatorname{rank}(\boldsymbol{B})=m$.

\section{Appendix C}
\begin{align*}
    n\cdot V_{AC,\hat \beta_1} = (1+\kappa)\left( \sum_{g=1}^p c_g \beta_g^2 + \sum_{g=1}^p\sum_{h=g+1}^p d_{gh}\beta_g\beta_h + Const\right)
 \end{align*}
 where
 \begin{align*}
c_g&=\sum_{\substack{j=1\\j\ne g}}^p r_{1j}^2\left( \sigma_{gj}^2+\sigma_g^2 \sigma_j^2\right)\left( \frac{1}{q_{gj}}-\frac{1}{q_j} \right)\\
    &\qquad+\frac{\kappa}{1+\kappa}\sum_{\substack{j=1\\j\ne g}}^p r_{1j}^2\sigma_{gj}^2\left( \frac{1}{q_{gj}}-\frac{1}{q_j} \right)\\
    &\qquad+\sum_{\substack{j=1 \\ j\ne g}}^p \sum_{\substack{k=j+1\\k\ne g}}^p 2 r_{1j}r_{1k}\left( \sigma_{gj}\sigma_{gk}+\sigma^2_g\sigma_{jk} \right)\left(\frac{q_{jk}}{q_jq_k}+\frac{q_{gjk}}{q_{gj}q_{gk}}-\frac{q_{gjk}}{q_jq_{gk}}-\frac{q_{gjk}}{q_kq_{gj}}\right)\\
     &\qquad+\frac{\kappa}{1+\kappa}\sum_{\substack{j=1 \\ j\ne g}}^p \sum_{\substack{k=j+1\\k\ne g}}^p 2 r_{1j}r_{1k}\sigma_{gj}\sigma_{gk}\left(\frac{q_{jk}}{q_jq_k}+\frac{q_{gjk}}{q_{gj}q_{gk}}-\frac{q_{gjk}}{q_jq_{gk}}-\frac{q_{gjk}}{q_kq_{gj}}\right)\\
d_{gh} & = \sum_{j=1}^p\sum_{k=j+1}^p 2 r_{1j}r_{1k}\sigma_{gk}\sigma_{hj}\left( \frac{q_{jk}}{q_jq_k} +\frac{q_{ghjk}}{q_{gj}q_{hk}}-\frac{q_{gjk}}{q_{k}q_{gj}}-\frac{q_{hjk}}{q_jq_{hk}}\right)\\
     &\qquad+  \sum_{j=1}^p\sum_{k=j+1}^p 2 r_{1j}r_{1k}\sigma_{gj}\sigma_{hk}\left( \frac{q_{jk}}{q_jq_k} +\frac{q_{ghjk}}{q_{gk}q_{hj}}-\frac{q_{gjk}}{q_jq_{gk}}-\frac{q_{hjk}}{q_kq_{hj}}\right)\\
     &\qquad+ \frac{\kappa}{1+\kappa} \mathop{\sum_{j=1}^p\sum_{k=j+1}^p}_{\{j=h\text{ or }k=g\}} 2 r_{1j}r_{1k}\sigma_{gj}\sigma_{hk}\left( \frac{q_{jk}}{q_jq_k} +\frac{q_{ghjk}}{q_{gk}q_{hj}}-\frac{q_{gjk}}{q_{j}q_{gk}}-\frac{q_{hjk}}{q_kq_{hj}}\right)\\
     &\qquad+ \frac{\kappa}{1+\kappa} \mathop{\sum_{j=1}^p\sum_{k=j+1}^p}_{\{j=g\text{ or }k=h\}} 2 r_{1j}r_{1k}\sigma_{gk}\sigma_{hj}\left( \frac{q_{jk}}{q_jq_k} +\frac{q_{ghjk}}{q_{gj}q_{hk}}-\frac{q_{gjk}}{q_{k}q_{gj}}-\frac{q_{hjk}}{q_jq_{hk}}\right)\\
     &\qquad+ \frac{\kappa}{1+\kappa} \mathop{\sum_{j=1}^p\sum_{j=i+1}^p}_{\{j\ne g,h;\;k\ne g,h\}} 2 r_{1j}r_{1k}\sigma_{gk}\sigma_{hj}\left( \frac{q_{jk}}{q_{j}q_{k}} +\frac{q_{ghjk}}{q_{gk}q_{hj}}-\frac{q_{gjk}}{q_{j}q_{gk}}-\frac{q_{hjk}}{q_kq_{hj}}\right)\\
     &\qquad+ \frac{\kappa}{1+\kappa} \mathop{\sum_{j=1}^p\sum_{k=j+1}^p}_{\{j\ne g,h;\;k\ne g,h\}} 2 r_{1j}r_{1k}\sigma_{gj}\sigma_{hk}\left( \frac{q_{jk}}{q_{j}q_{k}} +\frac{q_{ghjk}}{q_{gj}q_{hk}}-\frac{q_{gjk}}{q_{k}q_{gj}}-\frac{q_{hjk}}{q_jq_{hk}}\right)\\
     &\qquad+  \mathop{\sum_{j=1}^p\sum_{k=j+1}^p}_{\{j\ne g,h;\;k\ne g,h\}} 2 r_{1j}r_{1k}\sigma_{gh}\sigma_{jk}\left( \frac{2q_{jk}}{q_jq_k}+\frac{q_{ghjk}}{q_{gj}q_{hk}}+\frac{q_{ghjk}}{q_{gk}q_{hj}}-\frac{q_{gjk}}{q_jq_{gk}}-\frac{q_{gjk}}{q_{k}q_{gj}} -\frac{q_{hjk}}{q_jq_{hk}}-\frac{q_{hjk}}{q_kq_{hj}}\right)\\
     &\qquad+  \sum_{\mathclap{\substack{j=1\\j\ne g,h}}}^p 2r_{1j}^2\left( \sigma_{gj}\sigma_{hj}+\sigma_{gh}\sigma_j^2\right)\left( \frac{q_{ghj}}{q_{gj}q_{hj}} -\frac{1}{q_j}\right)\\
     &\qquad+ \frac{\kappa}{1+\kappa} \sum_{\mathclap{\substack{j=1\\j\ne g,h}}}^p 2r_{1j}^2\sigma_{gj}\sigma_{hj}\left( \frac{q_{ghj}}{q_{gj}q_{hj}} -\frac{1}{q_j}\right)\\
Const &= \left(\sum_{j=1}^p \frac{r_{1j}^2 \sigma_j^2 }{q_j} + \sum_{j=1}^p\sum_{k=j+1}^p 2 r_{1j}r_{1k}\sigma_{jk}\frac{q_{jk}}{q_jq_k}\right)\sigma^2.
 \end{align*}
In addition, 
 \begin{align*}
    n\cdot V_{CC,\hat \beta_1} =(1+\kappa) \left(\sum_{j=1}^p r_{1j}^2 \sigma_j^2  + \sum_{j=1}^p\sum_{k=j+1}^p 2 r_{1j}r_{1k}\sigma_{jk}\right)\frac{\sigma^2}{\tilde q} = (1+\kappa)\frac{r_{11}\sigma^2}{\tilde q}.
 \end{align*}

\section{Appendix D}
There several ways to estimate the kurtosis parameter $\kappa$ from the dataset.

\noindent 1) According to \citet{mardia1970measures} and \citet{wesselman1987elliptical}:
\begin{align*}
    \hat \kappa = \frac{\vec{\Sb^{-1}}\, \hat \Pi\, \vec{\Sb^{-1}}}{(p+3)(p+1)}-1,
\end{align*}
where $\hat \Pi$ is the $((p+1)^2\times (p+1)^2)$ matrix that contains all fourth-order central moment estimators. Then in \citet{maruyama2003estimation}, they used following expression:
\begin{align*}
    \hat \kappa = \frac{\sum_{i=1}^n \left\{ (\Zb_{\cdot i}-\bar \Zb)^T \Sb^{-1} (\Zb_{\cdot i}-\bar \Zb)\right\}^2}{n(p+3)(p+1)}-1,
\end{align*}
where $\Zb_{\cdot i}$ is the vector of observation of the $i$th sample; $\bar \Zb$ is sample mean of vector $\Zb$.\\

\noindent 2) We calculate the kurtosis parameter based on the estimated kurtosis of marginal random variable $Z_j$ \citep{wesselman1987elliptical}. 
\begin{align*}
    \hat \kappa = \frac{\sum_{i=j}^{p+1}\hat \kappa_j}{3(p+1)},
\end{align*}
where $\hat \kappa_j$ the corrected sample excess kurtosis of $Z_j$:
\begin{align*}
    \hat\kappa_j &= \frac{n-1}{(n-2)(n-3)}\left\{(n+1)\kappa_{j0}-3(n-1)\right\},\\
    \kappa_{j0}&=\frac{\frac{1}{n}\sum_{i=1}^n (Z_{ji}-\bar Z_j)^4}{\left\{\frac{1}{n}\sum_{i=1}^n(Z_{ji}-\bar Z_j)^2\right\}^2}.
\end{align*}

\section{Appendix E}
\noindent\textbf{Scenario 1, p=2:}\\
The sufficient condition that the CC estimator of $\beta_1$ is asymptotically better than the AC is $(1+\kappa)\left(\frac{1}{q_{-1}}-\frac{1}{q_1}\right)\left(2r_{11}-r_{11}^2\sigma_1^2\right)\sigma^2<0$, which is equivalent to:
\begin{align*}
    2r_{11}-r_{11}^2\sigma_1^2<0.
\end{align*}
We know $r_{11}$ is the top left element of $\sb^{-1}$, so that $r_{11}=\frac{\sigma_2^2}{\sigma_1^2\sigma_2^2-\sigma_{12}^2}$. Plus this into the inequality and immediately obtain $ \frac{\sigma_1^2\sigma_2^2}{\sigma_{12}^2}<2$.\\

\noindent\textbf{Scenario 2, $\bf p\ge3$:}\\
The sufficient condition that the CC estimator of $\beta_1$ is asymptotically better than the AC is $(1+\kappa)\left(\frac{1}{q_{-1}}-\frac{1}{q_1}\right)\left(2r_{11}-r_{11}^2\sigma_1^2\right)\sigma^2<0$, which is equivalent to:
\begin{align*}
    2r_{11}-r_{11}^2\sigma_1^2<0.
\end{align*}
We know $r_{11}$ is the top left element of $\sb^{-1}$, so that $r_{11}=\frac{(p-2)\sigma_{2'3'}+\sigma_{2'}^2}{-(p-1)\sigma_{12'}^2+(p-2)\sigma_1^2\sigma_{2'3'}+\sigma_1^2\sigma_{2'}^2}$. Plus this into the inequality and immediately obtain $\{(p-2)\sigma_{2'3'}+\sigma_{2'}^2\}\sigma_1^2<2(p-1)\sigma_{12'}^2$.\\

\section{Appendix F}
Function $f(\bb)$ is as following:
\begin{align*}
   f(\bb)=& -r_{11}^2(\sigma_{12'}^2+2\kappa \sigma_{12'}^2+\sigma_1^2\sigma_{2'}^2+\kappa \sigma_1^2\sigma_{2'}^2)\left(\frac{1}{q_{-1}}-\frac{1}{q_1}\right) \sum_{g=2}^p\beta_g^2\\
    &\qquad - 2r_{11}^2(\sigma^2_{12'}+2\kappa\sigma_{12'}^2+\sigma_1^2\sigma_{2'3'}+\kappa\sigma_1^2\sigma_{2'3'})\left( \frac{1}{q_{-1}}-\frac{1}{q_1} \right) \sum_{g=2}^p\sum_{h=g+1}^p \beta_g\beta_h \\
    & \qquad+ (1+\kappa)\left(\frac{1}{q_{-1}}-\frac{1}{q_1}\right)\left(2r_{11}-r_{11}^2\sigma_1^2\right)\sigma^2.
\end{align*}
We first show $\bb=0$ is the unique maximum point of $f(\bb)$. Let $A=-r_{11}^2(\sigma_{12'}^2+2\kappa \sigma_{12'}^2+\sigma_1^2\sigma_{2'}^2+\kappa \sigma_1^2\sigma_{2'}^2)\left(\frac{1}{q_{-1}}-\frac{1}{q_1}\right)$, $B=- 2r_{11}^2(\sigma^2_{12'}+2\kappa\sigma_{12'}^2+\sigma_1^2\sigma_{2'3'}+\kappa\sigma_1^2\sigma_{2'3'})\left( \frac{1}{q_{-1}}-\frac{1}{q_1} \right) $, then the Jacobian matrix of $f(\bb)$ is:
\begin{align*}
    f'(\bb)=\begin{pmatrix}
        2A\beta_2+B \sum_{\substack{j=1 \\ j\ne 2
        }}^p\beta_j\\
        2A\beta_3+B \sum_{\substack{j=1 \\ j\ne 3
        }}^p\beta_j\\
        \vdots\\
        2A\beta_p+B \sum_{\substack{j=1 \\ j\ne p
        }}^p\beta_j
    \end{pmatrix}^T.
\end{align*}
The Hessian matrix is:
\begin{align*}
    f''(\bb)&=\begin{pmatrix}
        2A & B & \cdots &B\\
        B & 2A  & \cdots &B\\
        \vdots &&& \vdots\\
        B &B& \cdots& 2A
    \end{pmatrix}.
\end{align*}
The $k$th order leading principal miner of $f''(\bb)$ is $(2A-B)^{k-1}\left(2A+(k-1)B\right)$. We can prove that
\begin{align*}
    2A-B &=2r_{11}^2\left( \frac{1}{q_{-1}}-\frac{1}{q_1} \right) \left(\sigma_1^2\sigma_{2'3'}+\kappa \sigma_1^2\sigma_{2'3'}-\sigma_1^2\sigma_{2'}^2-\kappa \sigma_1^2\sigma_{2'}^2\right)\\
    &=2r_{11}^2\left( \frac{1}{q_{-1}}-\frac{1}{q_1} \right) \sigma_1^2 (1+\kappa)(\sigma_{2'3'}-\sigma_{2'}^2)<0.\\
    2A+(k-1)B  &= -2r_{11}^2\left( \frac{1}{q_{-1}}-\frac{1}{q_1} \right) \left[k(1+2\kappa)\sigma_{12'}^2+(1+\kappa)\sigma_1^2\sigma_{2'}^2+(1+\kappa)(k-1)\sigma_1^2\sigma_{2'3'}\right]\\
    &= -2r_{11}^2\left( \frac{1}{q_{-1}}-\frac{1}{q_1} \right) \Big[(1+\kappa)\big\{-k\sigma_{12'}^2+\sigma_1^2\sigma_{2'}^2+(k-1)\sigma_1^2\sigma_{2'3'}\big\} \\
    &\qquad\qquad\qquad\qquad\qquad\qquad  +k(2+3\kappa)\sigma_{12'}^2 \Big]<0 \text{  (By C 1.1 in Appendix F)}.
\end{align*}
Therefore, we conclude that $f''(\bb)$ is negative definite by Sylvester's criterion and $\bb=\boldsymbol{0}$ is the unique maximum point of $f(\bb)$. 

The maximum value of $f(\bb)$ is $(1+\kappa)\left(\frac{1}{q_{-1}}-\frac{1}{q_1}\right)\left(2r_{11}-r_{11}^2\sigma_1^2\right)\sigma^2$, which is the constant term. Therefore, we claim that if the constant term (that does not involve $\bb$) in $f(\bb)$ is negative, $f(\bb)$ is always negative.

\section{Appendix G}
There are several constraints between $p$, $\kappa$, $\sigma_{2'3'}$, $\sigma_{2'}$, $\sigma_{12'}$, $\sigma_{1}$:

\noindent \textbf{C 1:} $\sx$ is positive definite, which is equivalent to its leading principal minors are positive. 

\textbf{C 1.1:} $(\sigma_{2'}^2-\sigma_{2'3'})^{p-2}\left(-(p-1)\sigma_{12'}^2+(p-2)\sigma_1^2\sigma_{2'3'}+\sigma_1^2\sigma_{2'}^2\right)>0$.

\noindent \textbf{C 2:} The ellipsoid is not degenerated (i.e., $\frac{2}{r_{11}}-\sigma_1^2>0$).

\textbf{C 2.1:} $(p-2)\sigma_{2'3'}\sigma_1^2+\sigma_1^2\sigma_{2'}^2>2(p-1)\sigma_{12'}^2$.

\noindent \textbf{C 3:} To make the elliptical distribution valid, $\kappa > -\frac{1}{2}$ \citep{bentler1986greatest}.\\
Therefore, we obtain the range of each parameter by solving all these conditions and thus get $I_L, I_R$.

The expressions of $M_1, M_2, M_3, M_4$ are:
 \begin{align*}
    M_1=&\frac{\sqrt{2\sigma_{12'}^2\left(p-1\right)\left((p-3)\sigma_{1}^2\sigma_{2'}^2+(2p-2+(p-2)(2+4\kappa)/(1+\kappa))\sigma_{12'}^2\right)}}{\sigma_{1}^2(p-2)}-\frac{\sigma_{1}^2\sigma_{2'}^2-2(p-1)\sigma_{12'}^2}{\sigma_{1}^2(p-2)}\\
    M_2=&\frac{1}{\sigma_{1}^2}\left(\sqrt{2\sigma_{12'}^2(p-1)\left((2+2p-2/(1+\kappa))\sigma_{12'}^2+(3-p)\sigma_{1}^2\sigma_{2'3'}\right)}+(2p-2)\sigma_{12'}^2\right)+(2-p)\sigma_{2'3'}\\
    M_3&=-\frac{(p-1)\sqrt{2\sigma_{12'}^2(2\sigma_{12'}^2-\sigma_{1}^2\sigma_{2'}^2)}}{\sigma_{1}^2(p-2)}-\frac{\sigma_{1}^2\sigma_{2'}^2-2(p-1)\sigma_{12'}^2}{\sigma_{1}^2(p-2)}\\
    M_4&=\frac{p-1}{\sigma_{1}^2}\left(\sqrt{2\sigma_{12'}^2(2\sigma_{12'}^2-\sigma_{1}^2\sigma_{2'3'})}+2\sigma_{12'}^2\right)+(2-p)\sigma_{2'3'}.
\end{align*}

\section{Appendix H}
\textbf{Proof of $V_{D,\hat\beta_1}\le 0$:}\\
The difference $V_{D,\hat\beta_1}$ is as follows:
\begin{align*}
    n\cdot V_{D,\hat\beta_1} 
    & = \left(\frac{1}{q_1}-\frac{1}{q_{-1}}\right)\left\{n(r_{11}-r_{11}^2\sigma_1^2)\sigma^2-c_1\beta_1^2\right\}.
\end{align*}
Then we prove $r_{11}-r_{11}^2\sigma_1^2\le 0$ and $c_1\ge 0$ so that $V_{D,\hat\beta_1}\le 0$.\\

\noindent 1. If we can prove $r_{11}\sigma_1^2\ge 1$, then it is obvious that $r_{11}-r_{11}^2\sigma_1^2\le 0$ holds. Since $\sb$ is symmetric and positive definite, then there exists a symmetric positive definite matrix $A$ such that $\sb=A^2$. Note that $\sigma^2_1=e^TA^TAe=\|Ae\|^2$ and $r_{11}=\|A^{-1}e\|^2$, 
where $e=(1,0,\cdots,0)^T\in \mathbb{R}^{p+1}$. By Cauchy-Schwarz, 
\begin{align}
    \sigma^2_1r_{11}=\|Ae\|^2\|A^{-1}e\|^2 \ge \langle Ae,A^{-1}e \rangle^2=e^T(A^{-1})^TAe=1\label{eqn:inq}.
\end{align}

\noindent 2. Now we show $c_1\ge 0$:
\begin{align*}
    c_1&=\sum_{\substack{j=2}}^p r_{1j}^2\left( \sigma_{1j}^2+\sigma_1^2 \sigma_j^2\right)
    +\kappa r_{1j}^2\sigma_1^2 \sigma_j^2
    +2\kappa r_{1j}^2\sigma_{1j}^2\\
    &\qquad \quad+\sum_{\substack{j=2}}^p \sum_{\substack{k=j+1}}^p 2r_{1j}r_{1k}\left( \sigma_{1j}\sigma_{1k}+\sigma^2_1\sigma_{jk} \right) 
    +2\kappa r_{1j}r_{1k}\sigma^2_1\sigma_{jk}
    + 4\kappa r_{1j}r_{1k}\sigma_{1j}\sigma_{1k}\\
    &=(1+\kappa)\sigma_1^2\left( \sum_{j=2}^pr_{1j}^2\sigma_j^2+\sum_{j=2}^p\sum_{k=j+1}^p2r_{1j}r_{1k}\sigma_{jk} \right)\\
    &\qquad \quad + (1+2\kappa)\left( \sum_{j=2}^pr_{1j}^2\sigma_{1j}^2+\sum_{j=2}^p\sum_{k=j+1}^p2r_{1j}r_{1k}\sigma_{1j}\sigma_{1k} \right)\\
    &=(1+\kappa)\sigma_1^2(r_{11}^2\sigma_1^2-r_{11})+(1+2\kappa)(1-r_{11}\sigma_1^2)^2\ge 0.
\end{align*}\\

\noindent\textbf{Proof of sufficiency (Proposition 3.5):}\\
\noindent In missing pattern (b), if the AC and the CC have the same asymptotic performance, then $X_1$ is independent with other predictors.\\
\noindent\textbf{Proof:}\\
When $V_{D,\hat\beta_1}=0$, we have $\sigma_1^2r_{11}=1$. By \eqref{eqn:inq}, $Ae$ should be linear dependent with $A^{-1}e$. Assuming $Ae=\lambda A^{-1}e\, (\lambda\ne 0)$, then we have $\lambda e=\sb e$, which implies $\sigma_{1j}=0\,(j\ge 2)$.\\

\noindent\textbf{Proof of necessity (Proposition 3.5):}\\
\noindent In missing pattern (b), if $X_1$ is independent with other predictors, then the AC and the CC have the same asymptotic performance.\\
\noindent\textbf{Proof:}\\
When $X_1$ is independent with other predictors (i.e., $\sigma_{1j}=0\,(j\ge 2)$), we can calculate $r_{11} = 1/\sigma_1^2$ and thus $r_{11}-r_{11}^2\sigma_1^2=0$. In addition, we have $r_{1j}=0\, (j\ge 2)$ and plug it into $c_1$:
\begin{align*}
    c_1&=\sum_{\substack{j=2}}^p r_{1j}^2\left( \sigma_{1j}^2+\sigma_1^2 \sigma_j^2\right)
    +\kappa r_{1j}^2\sigma_1^2 \sigma_j^2
    +2\kappa r_{1j}^2\sigma_{1j}^2\\
    &\qquad \quad+\sum_{\substack{j=2}}^p \sum_{\substack{k=j+1}}^p 2r_{1j}r_{1k}\left( \sigma_{1j}\sigma_{1k}+\sigma^2_1\sigma_{jk} \right) 
    +2\kappa r_{1j}r_{1k}\sigma^2_1\sigma_{jk}
    + 4\kappa r_{1j}r_{1k}\sigma_{1j}\sigma_{1k} = 0.
\end{align*}
Therefore, $V_{D,\hat\beta_1}=0$ in this condition. 

\clearpage
\section*{Table}
\setcounter{table}{0}

\newcommand\marktopleft[3]{
    \tikz[overlay,remember picture] 
        \node (marker-#1-a) at (#2,#3) {};
}
\newcommand\markbottomright[5]{
    \tikz[overlay,remember picture] 
        \node (marker-#1-b) at (#4,#5) {};
    \tikz[overlay,remember picture,#3,line width=0.2mm]
        \node[draw=#2,rectangle,fit=(marker-#1-a.center) (marker-#1-b.center)] {};%
}
\begin{table}[!h]
    \newcolumntype{Y}{>{\centering\arraybackslash $}X<{$}}
    \caption{Different covariance structures}\label{tab:missing_s}
    \begin{tabularx}{\textwidth}{>{$}c<{$}>{$}c<{$}>{$}c<{$}>{$}c<{$}YYYY}
        \toprule
         \rho_{VX} &\sigma_V^2 & \sigma_X^2 & \sigma_{VX} & \var(\hat\beta_{V,CC}^{\text{(s)}})^a & \var(\hat\beta_{V,CC}^{\text{(t)}})^a  & \var(\hat\beta_{V,AC}^{\text{(s)}})^a & \var(\hat\beta_{V,AC}^{\text{(t)}})^a \\
        \midrule
        \marktopleft{c1}{1pt}{8pt} 0.900 & 1.000 & 1.000 & 0.900 & \textbf{5.7814} & \textbf{5.8480} & 8.1862 & 8.1899 \\
        0.700 & 1.000 & 1.000 & 0.700 & \textbf{2.1508} & \textbf{2.1786} & 2.2062 & 2.2197 \\
        0.500 & 1.000 & 1.000 & 0.500 & 1.4629 & 1.4815 & \textbf{1.3955} & \textbf{1.4019} \\
        0.300 & 1.000 & 1.000 & 0.300 & 1.2070 & 1.2210 & \textbf{1.1193} & \textbf{1.1224} \\
        0.100 & 1.000 & 1.000 & 0.100 & 1.1112 & 1.1223 & \textbf{1.0196} & \textbf{1.0201} \\        
        \hdashline
        0.900 & 1.000 & 0.329 & 0.516 & \textbf{5.7598} & \textbf{5.8261} & 7.8642 & 7.8574 \\
        0.700 & 1.000 & 0.543 & 0.516 & \textbf{2.1523} & \textbf{2.1801} & 2.1854 & 2.1988 \\
        0.500 & 1.000 & 1.065 & 0.516 & 1.4629 & 1.4815 & \textbf{1.3968} & \textbf{1.4032} \\
        0.300 & 1.000 & 2.958 & 0.516 & 1.2070 & 1.2210 & \textbf{1.1433} & \textbf{1.1447} \\
        0.100 & 1.000 & 26.626 & 0.516 & \textbf{1.1112} & \textbf{1.1223} & 1.2595 & 1.2485 \\               
        \hdashline
        0.900 & 0.329 & 1.000 & 0.516 & \textbf{17.5069} & \textbf{17.7086} & 24.7473 & 24.7602 \\
        0.700 & 0.543 & 1.000 & 0.516 & \textbf{3.9637} & \textbf{4.0149} & 4.0663 & 4.0911 \\
        0.500 & 1.065 & 1.000 & 0.516 & 1.3736 & 1.3911 & \textbf{1.3103} & \textbf{1.3164} \\
        0.300 & 2.958 & 1.000 & 0.516 & 0.4080 & 0.4128 & \textbf{0.3784} & \textbf{0.3794} \\
        0.100\markbottomright{c1}{red}{solid}{-1pt}{-0pt} & 26.626 & 1.000 & 0.516 & 0.0417 & 0.0422 & \textbf{0.0383} & \textbf{0.0383}  \\
        \bottomrule       
    \end{tabularx}\\
    \footnotesize
    $^a$ (s) means simulation results; (t) means theoretical results using expressions~(1), (2) in Section 3.1. All the variances are in the order of magnitude of $-3$.\\
    Note: the smaller variances in each setting are in bold face.
\end{table}
\begin{table}[!h]
    \newcolumntype{Y}{>{\centering\arraybackslash $}X<{$}}
    \caption{Different residual variance}\label{tab:missing_r}
    \begin{tabularx}{\textwidth}{>{$}c<{$}YYYY}
        \toprule
        \sigma_\varepsilon^2 & \var(\hat\beta_{V,CC}^{\text{(s)}})^a & \var(\hat\beta_{V,CC}^{\text{(t)}})^a  & \var(\hat\beta_{V,AC}^{\text{(s)}})^a & \var(\hat\beta_{V,AC}^{\text{(t)}})^a \\
        \midrule
        0.010 & \textbf{0.0150} & \textbf{0.0151}  & 0.0351 & 0.0345 \\
        0.050 & 0.7476 & 0.7572 & \textbf{0.7263} & \textbf{0.7293} \\
        1.000 & 1.4952 & 1.5143  & \textbf{1.4313} & \textbf{1.4382} \\
        5.000 & 7.4762 & 7.5715  & \textbf{7.0710} & \textbf{7.1095} \\        
        \bottomrule       
    \end{tabularx}\\
    \footnotesize
    $^a$ (s) means simulation results; (t) means theoretical results using expressions~(1), (2) in Section 3.1. All the variances are in the order of magnitude of $-3$.\\
    Note: the smaller variances in each setting are in bold face.
\end{table}
\begin{table}[!h]
    \newcolumntype{Y}{>{\centering\arraybackslash $}X<{$}}
    \caption{Different true coefficients}\label{tab:missing_c}
    \begin{tabularx}{\textwidth}{>{$}c<{$}>{$}c<{$}YYYY}
        \toprule
        \beta_V & \beta_X & \var(\hat\beta_{V,CC}^{\text{(s)}})^a & \var(\hat\beta_{V,CC}^{\text{(t)}})^a  & \var(\hat\beta_{V,AC}^{\text{(s)}})^a & \var(\hat\beta_{V,AC}^{\text{(t)}})^a\\
        \midrule
        0.310 & 0.279 & 1.4952 & 1.5143  & \textbf{1.4313} & \textbf{1.4382} \\
        0.620 & 0.279 & 1.4952 & 1.5143  & \textbf{1.4313} & \textbf{1.4382} \\
        0.930 & 0.279 & 1.4952 & 1.5143 & \textbf{1.4313} & \textbf{1.4382} \\
        0.310 & 0.558 & 1.4952 & 1.5143  & \textbf{1.4949} & \textbf{1.4992} \\
        0.310 & 1.116 & \textbf{1.4952} & \textbf{1.5143}  & 1.7477 & 1.7433 \\        
        \bottomrule       
    \end{tabularx}\\
    \footnotesize
    $^a$ (s) means simulation results; (t) means theoretical results using expressions~(1), (2) in Section 3.1. All the variances are in the order of magnitude of $-3$.\\
    Note: the smaller variances in each setting are in bold face.
\end{table}

\clearpage
\bibliographystyle{asa}
\bibliography{arxiv}
\end{document}